\documentclass[%
 reprint,
 amsmath,amssymb,
 aps,
]{revtex4-2}

\usepackage{graphicx}
\usepackage{dcolumn}
\usepackage{bm}

\newcommand*{\rttensor}[1]{\overline{\overline{#1}}}
\usepackage{upgreek}

\begin{document}

\preprint{APS/123-QED}

\title{Active solid-state nanopores: Self-driven flows/chaos at liquid-gas nanofluidic interface}

\author{Vinitha Johny$^1$}
\author{Siddharth Ghosh$^{1,2,3}$}%
\email{Email of correspondence: sg915@cam.ac.uk}
\affiliation{%
$^1$International Center for Nanodevices, INCeNSE-TBI, Indian Institute of Science Campus, Bangalore 560 012, India.\\
$^2$International Center for Nanodevices, High Tech Campus Eindhoven, 5656 AE Eindhoven, The Netherlands.\\
$^3$Department of Applied Mathematics and Theoretical Physics, University of Cambridge, Cambridge CB3 0WA, UK.}%

\begin{abstract}
We present a study of self-driven flow dynamics at the liquid-gas interface within nanofluidic pores, devoid of any external driving forces. 
The investigation centres on the Rayleigh-Taylor instability phenomena occurring in sub-100 nanometre-scale fluidic pores situated within a micrometer-scale water and air domain. 
This research rigorously validates our flow velocity equation using simulation results while delving into the mass transfer efficiency of these intricate flow structures. Notably, we introduce a concept—an `active solid-state nanopore'—that exhibits self-driven flow switching behaviour, transitioning between active and passive states without the need for mechanical components. This study reveals highly nonlinear and complex fluid dynamics within nanoscale dimensions, marking an exploration in this domain at room temperature. Implications of self-driven nanofluidics extend across diverse fields, from enhancing biosensors and healthcare applications to advancing net-zero sustainable energy production and contributing to the fundamental understanding of fluid dynamics in confined spaces.
\end{abstract}

\maketitle

\section{Introduction}

Nanofluidic pores \cite{ali2017cesium, zhang2020engineering, ghosh2020single, zou2020transpiration}, which refer to sub-100 nm-scale conduits for fluid flow, have sought significant attention for electrical energy harvesting \cite{siria2013giant, Bocquet2017}. 
However, the exploration of mechanical energy extraction from nanofluidics remains relatively uncharted territory. Intriguingly, there is evidence that nanofluidic systems can generate substantial forces, exemplified by the flexible nanopores found in the stomata of leaves. These natural nanopores create remarkable negative pressure, on the order of megapascals, enabling them to transport water against gravity over substantial heights \cite{heydt1991measurement, zou2020transpiration}.
In this study, we delve into the concept of mechanical energy extraction from nanofluidics, specifically focusing on a transpirational approach \cite{gates1968transpiration, rashidi2014homotopy}.
This approach, inspired by nature, holds the promise of scalable designs suitable for very large scale integration/VLSI applications \cite{skaug2018nanofluidic}, enabling on-chip flow generation \cite{ahmed2020chip}, molecule sampling \cite{ruggeri2017single}, and nanorobotics \cite{cira2015vapour}. Furthermore, it aligns with the pursuit of clean energy solutions that offer precise fluid control at the scale of biological cellular nanopores, propelling us closer to the realm of artificial life and next-generation nanomechanically engineered transportation \cite{cheng1997three, de2001water, wang2011tunneling}.
To advance our understanding of interfacial fluid dynamics within nanometric cavities, we turn to the phenomenon of Rayleigh-Taylor instabilities. These instabilities emerge when two fluids of differing densities intersect \cite{sharp1984overview}, giving rise to intricate interference patterns at their interface due to molecular interactions \cite{linden1994molecular}. By examining the rate at which these instabilities develop, we gain valuable insights into the fluid dynamics of interfacial systems \cite{taylor1950instability}.
Remarkably, fluid dynamics in nanofluidic pores at the liquid-gas interface have remained relatively unexplored. This gap in research is not solely due to experimental challenges but also arises from theoretical complexities. 

\begin{figure}[h!]
\includegraphics[width=0.45\textwidth]{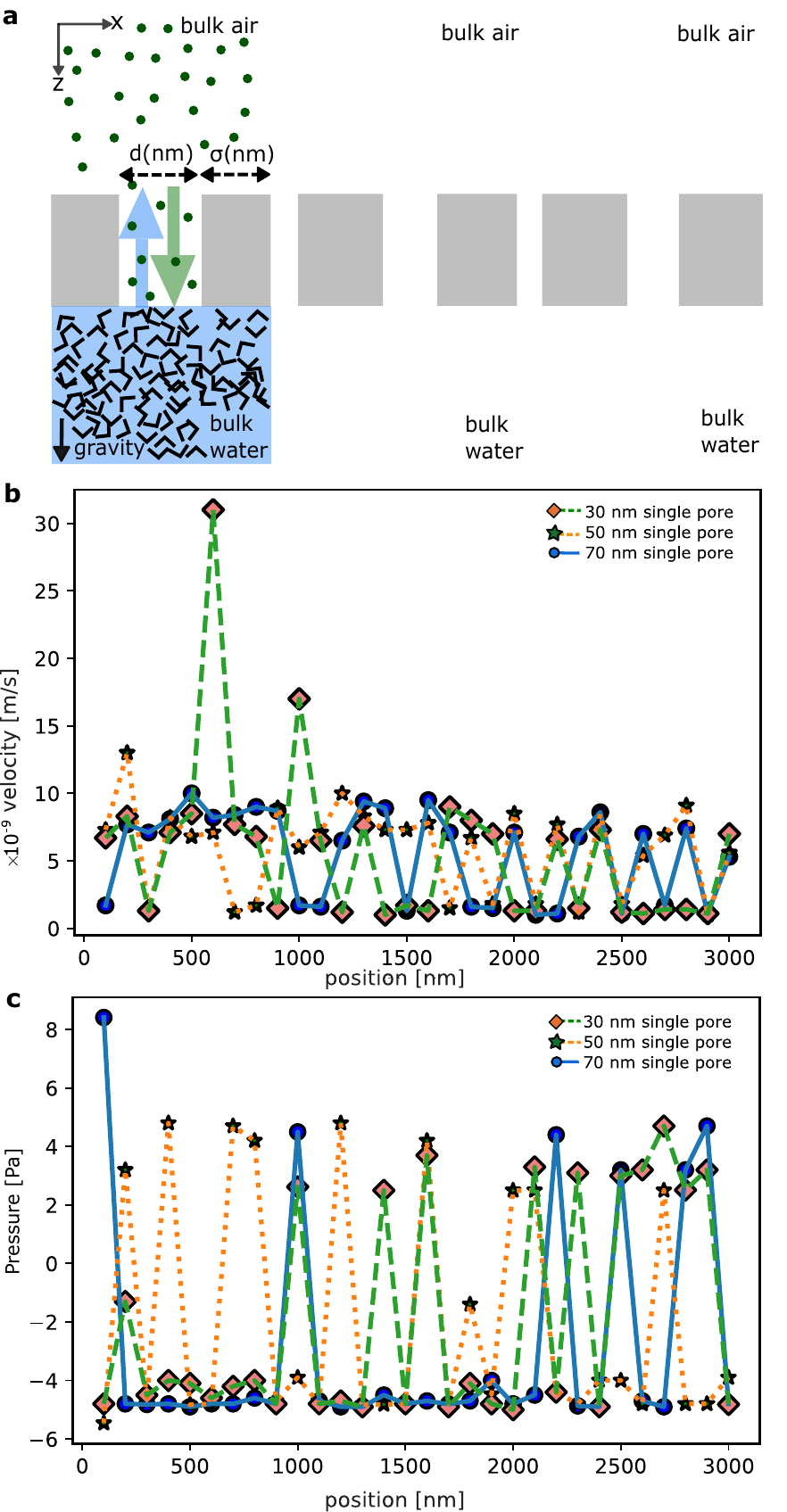}
\caption{
(a) Principle of self-driven flow at the interface of air and water through a single nanofluidic pore; $\sigma$(nm) is the distance of nanofluidic pore from the boundary, d(nm) is the nanofluidic pore throat diameter. 
(b) Chaotic oscillation of reference maximum velocity in mixed phase of water and air at the interface with single pore of d of 30 nm, 50 nm, and 70 nm. 
(c) Variation of pressure inside the nanofluidic due to the distance from edge, $\sigma$.}
\label{Figure:1}
\end{figure}

Herein, we present a novel study of self-driven flow dynamics through sub-100 nm nanofluidic pores, drawing inspiration from the behaviour of nanopores on leaf surfaces.
Our investigation employs a Rayleigh-Taylor instability model at the water-air interface to elucidate the nontrivial phenomenon of evaporation against gravity through nanofluidic pores, mirroring the mechanisms at play in natural leaf surfaces. Notably, we unveil the dynamics of flow in stable solid-state nanofluidic pores without the application of external voltage or pressure against gravity. While conventional nanofluidic transport mechanisms have primarily revolved around electrokinetic and pressure-driven methods \cite{meller2002single, xuan2007solute}, the transpirational approach, as observed in Figure \ref{Figure:1}a, introduces unique factors such as van der Waals forces, electrostatic interactions, induced dipoles between polar and non-polar species, and oscillatory solvation pressure. The results of our study reveal both linear and chaotic flow patterns in microreservoirs featuring nanofluidic pores, with significant variations in velocity across different pore positions.
The findings of this research not only uncover the exotic fluid dynamics governing sub-100 nm nanofluidic pores but also open up exciting possibilities for engineering self-driven nanopores using solid-state materials. This pioneering exploration prompts new questions about interfacial hydraulic resistance, which may find relevance in understanding phenomena such as stomatal closure in plants, and sparks innovative ideas for nanoengineered propulsion in transportation \cite{10.1093/forestscience/40.3.513}.

\begin{figure}[]
\centering
\includegraphics[width=0.4\textwidth]{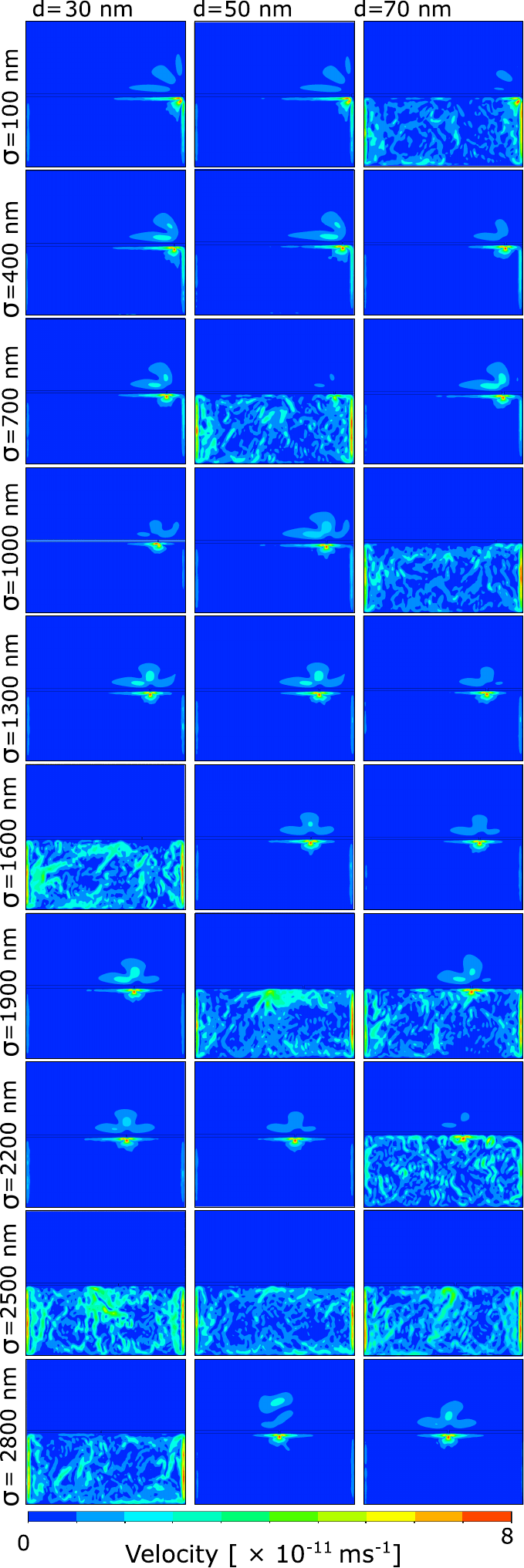}
\caption{\textbf{Pattern of flow dynamics in bulk water and bulk air due to single nanopores} by varying its diameter, $d$ and position from the boundary, $\sigma$ --- column 1, column 2, and column 3 are for diameters 30 nm, 50 nm, and 70 nm, respectively. In all cases,  $\sigma$ is varied from the 100 nm to 2800 nm referring position from the boundary to the centre of the systems.}
\label{Figure:2}
\end{figure}

\subsection*{Model Description}
The model investigates fluid behaviour at the nanofluidic liquid-gas interface between bulk water and bulk air. Figure \ref{Figure:1}a provides an overview of the system with unidirectional flow patterns.
To set the stage, we consider a fluidic reservoir measuring 6000 nm $\times$ 6000 nm, divided by a 100 nm wall. This silicon wall comprises 96\% of the separation at the water-air interface. Key material properties include a density of $2.18 \times 10^{-12}$ g/m$^3$, elastic modulus of $6.8 \times 10^{10}$ Pa, shear modulus of $2.8 \times 10^{10}$ Pa, Poisson's ratio of 0.19, thermal conductivity of 1.38 W/mK, and specific heat of 750 J/kg$^\circ$C. We utilise a triangular mesh geometry with an average element area of $9 \times 10^{-12}$ m$^{2}$ and an element mesh size of $2.5 \times 10^{-8}$ m. The viscous model is set to laminar, and implicit formulation is employed. The system temperature is maintained at 288.16 K.
Our model is based on the Navier-Stokes equations, which consist of the continuity, momentum, and energy equations. Additionally, we employ a volume fraction equation to account for multiphase flow. The general fluid transport equation is given as:
\begin{equation}
    \frac{\partial}{\partial t} (\gamma \rho \kappa) + \nabla.(\gamma \rho \bm{v} \kappa)= \nabla. {\rttensor\tau} + M\kappa
\end{equation}
Here, $\gamma$ represents the phase volume fraction, $\rho$ is the phase density, $\kappa$ is the phase variable, $\bm v$ is the phase velocity, $M$ is the source term for added mass transfer, and ${\rttensor\tau}$ is the diffusion term (stress tensor).
The volume fraction equation is described as:
\begin{equation}
    \frac{1}{\rho_l}\bigg[\frac{\partial}{\partial t}(\gamma_l \rho_l) + \nabla.(\gamma_l \rho_l \bm{v}_l)\bigg] = M \gamma_l + \sum_{p=1}^n ({\dot{m}}_{gl} - {\dot{m}}_{lg})
\end{equation}
This equation considers mass transfer through the nanofluidic pores, regardless of the flow direction.
To study the system's behaviour, we vary the diameter of single nanofluidic pores (\(d\)) at 30 nm, 50 nm, and 70 nm. We also vary the distance (\(\sigma\)) from the boundary of the reservoir in the orthogonal direction to gravity in 100 nm steps from 100 nm to 3000 nm at the center.
It is important to note that the mesoscopic physics of the fluidic interface at the nanometric length scale adds complexity to our study.
All simulations were performed in two dimensional and performed using finite element method Ansys Fluent 2021 R1 solver in Intel i7 CPU.

\subsection*{Single Nanofluidic Pore at Liquid-Gas Interface}
Although the physical model of the leaf-like system shown in Figure \ref{Figure:1}a appears geometrically simple, the fluid mechanics revealed non-intuitive behaviour that defies a linear trend. 
One might expect the maximum velocity of fluid flow to occur at the nanofluidic interface, resulting in a plateaued velocity profile. However, as shown in Figure \ref{Figure:1}b and Figure \ref{Figure:1}c, we observe strong fluctuations in reference velocities and pressure.
For a 30 nm nanofluidic pore, varying the pore's position by 200 nm produces velocity steps of 0 m/s, $4.5 \times 10^{-9}$ m/s, and $30 \times 10^{-9}$ m/s. Similarly, in a 50 nm nanofluidic pore system, we observe velocity steps of 0 m/s and $10 \times 10^{-9}$ m/s within a 100 nm distance. Surprisingly, there is no flow between 2500 nm and 2900 nm, followed by a sudden return of flow at 3000 nm. 
This strong oscillation nearly stabilises within the range of 900 nm to 1600 nm, with velocities ranging from $4.5 \times 10^{-9}$ m/s to $10 \times 10^{-9}$ m/s, only to continue with strong oscillations over distance. A similar pattern is observed in the 70 nm nanopore system within $\sigma =$ 100 nm. 
A nearly stable plateau with an average velocity of $7.5 \times 10^{-9}$ m/s can be observed within $\sigma =$ 200 nm to 900 nm, followed by continued strong fluctuations.
The aforementioned oscillations prompted us to investigate the overall flow patterns of the system. Figure \ref{Figure:2} illustrates the distribution of velocity magnitude over two spatial dimensions, \(x\) and \(y\). 
The most common flow pattern in single nanofluidic pore systems resembles a cloud or hat shape with large flow velocities. The movement of liquid from the water phase to the air phase encounters hindrance by back-flow, as depicted in Figure \ref{Figure:1}c, often resulting in the formation of a visible meniscus.
Interestingly, we observe the sudden collapse of flow in the air region at certain positions of nanofluidic pores, varying the distance from 100 nm to 3000 nm. A driving mechanism that initiates and inhibits flow appears to be continuous. 
The pressure drops above the meniscus induce back-flow in the water phase, initiating motion. This phenomenon is observed in several instances across different pore sizes and positions.
In Figure \ref{Figure:2}, we notice distinct small peaks of counter-flow on either side of the highest peak, similar to the Gibbs-Marangoni flow \cite{narayanan2011interfacial}.
The maximum flow velocity through 30 nm, 50 nm, and 70 nm nanofluidic pores are found to be approximately $1.3 \times 10^{-11}$ m/s, $1.3 \times 10^{-9}$ m/s, and $8.4 \times 10^{-9}$ m/s, respectively, when the reference flow velocity through the nanofluidic pore is set to be $4.2 \times 10^{-9}$ m/s.

\textit{Zero-flow nano-chaos or 0FNC state}: 
In several cases, we observed no flow in the air phase while the water phase exhibited significant motion with strong velocity components near the boundary. We have termed this phenomenon ``zero-flow nano-chaos" (0FNC state).
For instance, in Figure \ref{Figure:2}, 0FNC state is evident in the 30 nm nanofluidic pore when positioned at 1600 nm, 2500 nm, and 2800 nm from the right boundary. Appendix Figures 6 and 7 further illustrates discrete events of 0FNC state at various positions, such as 300 nm, 1200 nm, 2000 nm, 2300 nm, and a range from 1400 nm to 1600 nm, as well as from 2500 nm to 2900 nm.
While 50 nm and 70 nm nanofluidic pores do not consistently exhibit 0FNC state at the same positions as the 30 nm pore, they do display this behaviour at specific locations. Appendix Figure 6 demonstrates events of 0FNC state for the 50 nm nanofluidic pore at 700 nm, 1900 nm, and 2500 nm.
At $\sigma =$ 700 nm, we noted a transition with a small flow component in the air phase. At $\sigma =$ 2500 nm, 0FNC state was observed in all cases.
The direction of vortices and their pairs may have a certain relation with 0FNC state, a topic we discuss in the subsequent Discussion section.
Appendix Figure 6 reveals a range of positions within $\sigma =$ 2500 nm to 2900 nm where the 50 nm nanofluidic pore experiences three events of flow in the air phase, breaking the 0FNC state. 
Similarly, Appendix Figure 7 demonstrates a pattern similar to that of the 30 nm pore for the 70 nm pore, except at $\sigma =$ 2600 nm, where we found a transition from 0FNC state to flow in the air.

\begin{figure*}[]
\includegraphics[width =0.53\textwidth]{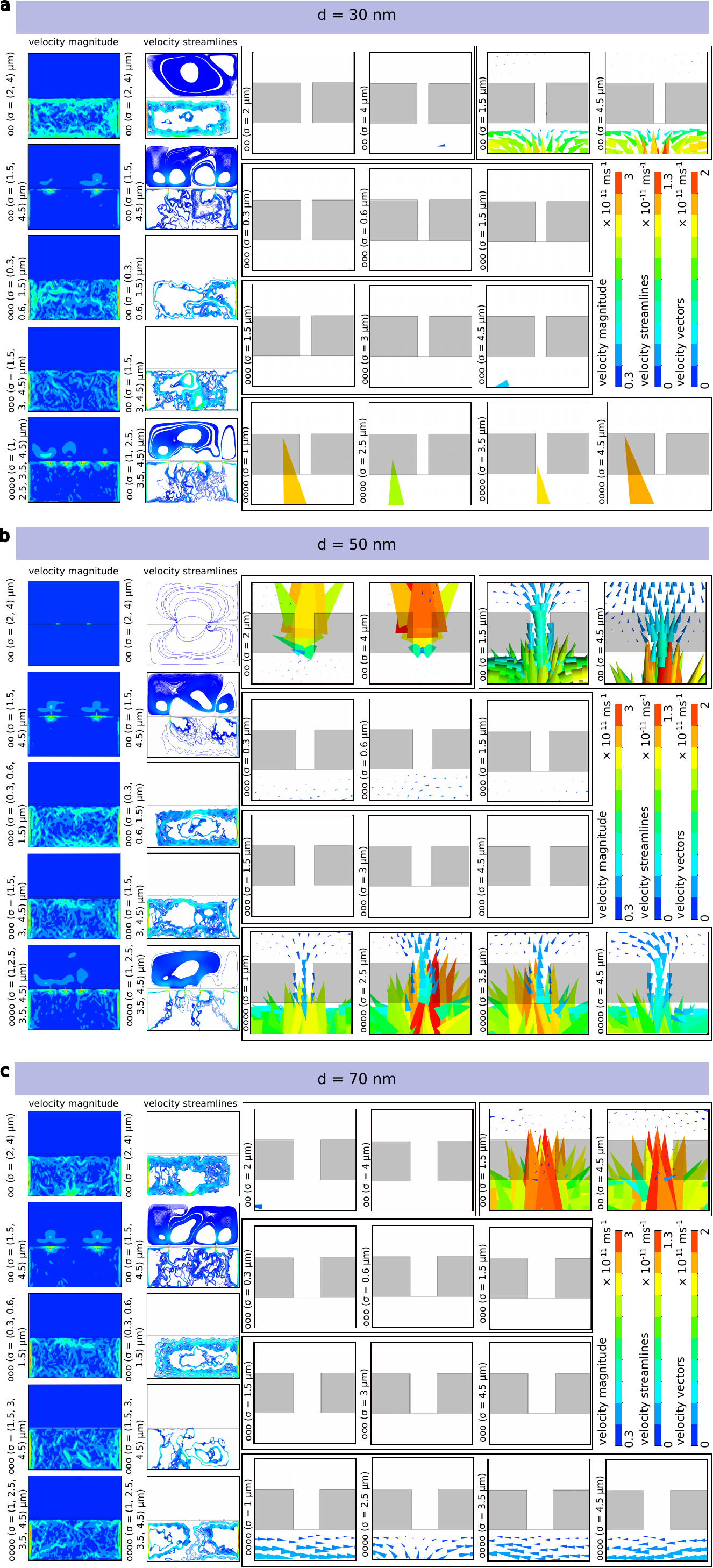}
\caption{\textbf{Flow patterns of multiple nanofluidic pore systems}.
Global/complete views of velocity flow pattern with velocity vectors in the 1st column and its corresponding velocity stream lines in the 2nd column, and followed by these global views, 4th to 7th columns represent local components of velocity vectors of local flow at the individual nanofluidic pores in the same order of their presence from left to right in the global view.
(a) 30 nm showing no prominent flow from one fluid phase to another in two pores (oo), three pores (ooo), and four pores (oooo), (b) 50 nm with two pores (oo) possessing strong flow from water to air as well as moderate flow from air to water, three pores (ooo) with no flow from different fluid phases, and four pores (oooo) showing weak flow components from one fluid phase to another, and (c) 70 nm with no flow component from one fluid phase to another in two pores (oo), three pores (ooo), and four pores (oooo)}.
\label{Figure:3}
\end{figure*}

\begin{figure*}[ht]
\includegraphics[width=1\textwidth]{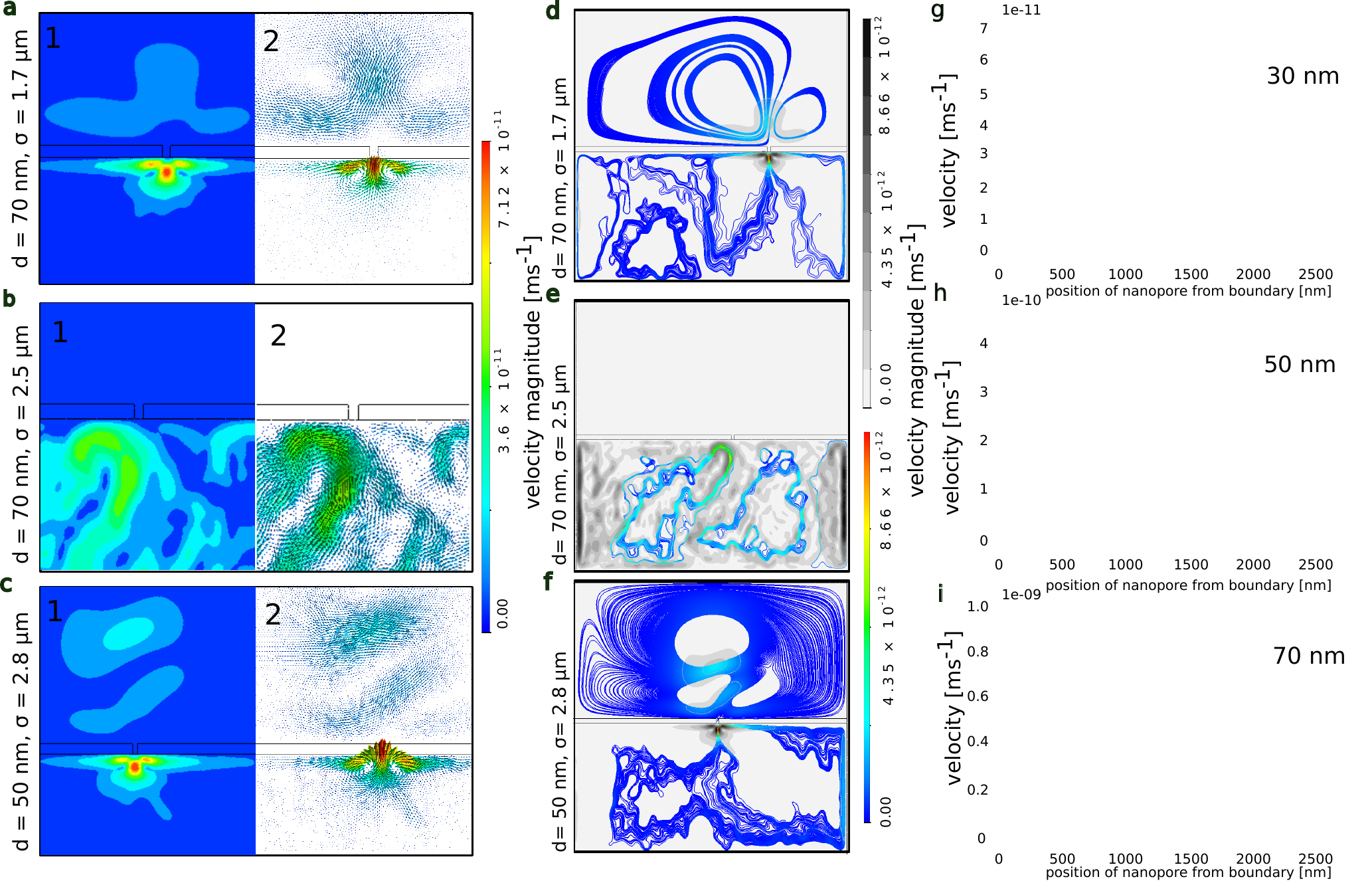}
\caption{ \textbf{Exemplary local flow patterns at the vicinity of nanofluidic pore.}
(a) Flow dynamics of single nanofluidic pore system with $d$ = 70 nm situated at $\sigma$ = 1700 nm; 
1 -- velocity magnitude distribution and 2 -- velocity vectors indicating vortices in water and air resulting with no flow from water to air. 
Have (b) Single nanofluidic pore system with $d$ = 70 nm situated at $\sigma$ = 2500 nm;
1 -- velocity magnitude with high value near the opening and 2 -- velocity vector with no flow in air but flow component.
(c) 50 nm single nanofluidic pore system at $\sigma$ = 2800 nm; 1 -- velocity magnitude and 2 -- velocity vector with prominent back flow and vertices in water and air.
\textbf{Global flow patterns and velocity streamlines in the systems for single nanopore.} (d) Complete view of (a) showing no connection of streamlines from water to air and vice versa. 
(e) Complete view of (b) with flow in air, significant chaotic flow patterns in air phase, and strong velocity magnitude at the boundary.
(f) Complete view of (c) with the largest velocity magnitude near nanofluidic pore in water and cavitation being formed in the air phase.  
\textbf{Analytical and simulated mass flow velocity through nanopore as function of $\sigma$.}
(g) $d$ = 30 nm with prominent errors for $\sigma$ at 400 nm and 700 nm
(h) $d$ = 50 nm with prominent errors for $\sigma$ at 100 nm, 700 nm, and 1300 nm,
(i) $d$ = 70 nm with prominent errors in all $\sigma$ except 1600 nm.
}
\label{Figure:4}
\end{figure*}

\subsection*{Multiple Nanofluidic Pores at Liquid-Gas Interface}
We have conducted an investigation into the behaviour of multiple nanofluidic pore systems at the liquid-gas interface. In Figure \ref{Figure:3}, we present the flow behaviour of these systems, varying the number of nanofluidic pores from two to four, while studying their spatial dependencies. 
Two cases of two-pore systems are examined based on pore positions: $\sigma = (2, 4)~\upmu$m and $\sigma = (1.5, 4.5)~\upmu$m. For three-pore systems, variations include $\sigma = (0.3, 0.6, 1.5)~\upmu$m and $\sigma = (1.5, 3, 4.5)~\upmu$m. In the case of four-pore systems, we focused on one scenario with $\sigma = (1, 2.5, 3.5, 4.5)~\upmu$m.
We provide global views of velocity flow patterns with velocity vectors in the first column and corresponding velocity streamlines in the second column for three nanofluidic pore sizes: 30 nm (Figure \ref{Figure:3}a), 50 nm (Figure \ref{Figure:3}b), and 70 nm (Figure \ref{Figure:3}c). 
Subsequently, we present local components of velocity vectors for each nanofluidic pore.
In two-pore systems, both flow and 0FNC state are observed (Figure \ref{Figure:3}a). 
For three-pore systems, 0FNC conditions persist. 
In the streamline visualisation, we observe clustered flow in the central region of the water phase. 
Magnified views of these pores do not show flow components inside the pore. For four-pore systems, flow is visible through all pores with clustered flow streamlines oriented towards the air phase from water to air phases.

\textit{30 nm Multipore Systems}: 
In two-pore systems (Figure \ref{Figure:3}a.oo), we observe both 0FNC state and fully-developed flow behaviour. Notably, when the pore spacing ($\sigma$) is set to (2, 4) µm, 0FNC predominates within the water phase. A closer inspection of pores at $\sigma = 2$ µm and $\sigma = 4$ µm (Figure \ref{Figure:3}a.oo) reveals the absence of flow at the water-air interface within the pores. However, by adjusting the pore positions closer to the nearest boundaries (by 500 nm) at $\sigma = (1.5, 4.5)$ µm, the system transitions into a state of flow, with flow rates of $2.4 \times 10^{-11}$ ms$^{-1}$ and $3 \times 10^{-11}$ ms$^{-1}$ through the first and second pores, respectively. The velocity streamlines indicate a synchronised global flow between the two pores. In the magnified views of pores at $\sigma = 1.5$ µm and $\sigma = 4.5$ µm, minor flow components are observed at the water-air interface within the pores, with significantly larger flow components near the pores in the water phase. 
Expanding to three-pore systems (Figure \ref{Figure:3}a.ooo), we again observe a 0FNC condition. Streamline visualisations at $\sigma = (1.5, 3.0, 4.5)$ µm show clustered flow in the central region of the water phase. However, magnified views of groups of three sets of pores in both cases do not display any flow components within the pores (e.g., Figure \ref{Figure:3}a.ooo at $\sigma = (0.3, 0.6, 1.5)$ µm and Figure \ref{Figure:3}a.ooo at $\sigma = (1.5, 3, 4.5)$ µm). 
Further increasing the number of pores to four (Figure \ref{Figure:3}a.oooo), flow is visible through all pores, with clustered flow streamlines oriented towards the pores and the air phase, transitioning from the water phase to the air phase. In the magnified view of these pores at $\sigma = (1, 2.5, 3.5, 4.5)$ µm, velocity vectors within each pore exhibit orientation towards the air phase (forward flow). Notably, the number of pores displaying forward flow vectors exceeds those with backward flow vectors (towards the water phase), explaining the higher total flow rate towards the air phase.

\textit{50 nm Multipore Systems:} 
When we increased the pore size to 50 nm, 0FNC state decreased. Flow components towards the air phase are observed when $\sigma = (2, 4)~\upmu$m. Changing the position of two pores towards the nearest boundaries ($\sigma = (1.5, 4)~\upmu$m) results in reversed flow towards the water phase through one pore and a 2$\pi$ transition towards the air phase through the other.
For three-pore systems, 0FNC state is prominent in the water phase. Magnified views of these pores do not show flow vectors inside the pores. However, in four-pore systems, continuous and prominent flow through all pores is observed.

As shown in Figure\ref{Figure:3}b, we investigate all the same positions of 30 nm systems with change in pore size to 50 nm.
The 0FNC systems have decreased as the pore size increased from 30 nm to 50 nm.
When $\sigma=~(2,~4)~\upmu$m, the flow components are significant towards the air phase. 
The magnified view of these pores as in Figure 3b.oo($\sigma=2~\upmu$m) and Figure 3b.oo($\sigma=4~\upmu$m) shows
a high magnitude stream of velocity vectors towards the air phase.
As the position of each nanopore is changed by 500 nm towards their nearest boundaries, the flow dynamics is observed to change considerably as in Figure \ref{Figure:3}b.oo($\sigma=~(1.5,~4)~\upmu$m.
In the magnified view of the pore in Figure \ref{Figure:3}b.oo($\sigma=~1.5~\upmu$m), the flow vectors are oriented towards the water phase.
Here, a reverse flow towards the water phase with magnitude of $1.5\times10^{-11}$ ms$^{-1}$ is observed.
In Figure \ref{Figure:3}b.oo($\sigma= 4~\upmu$m), the flow vectors takes a 2$\pi$ transition towards the air phase.
When the number of pores are increased to three, 0FNC is prominent in water phase as in Figure \ref{Figure:3}b.ooo.
The magnified pore views of three pore systems of pore size 50 nm in Figure \ref{Figure:3}b.ooo($\sigma=~(0.3,~0.6,~1.5~)~\upmu$m) and Figure \ref{Figure:3}b.ooo($\sigma=~(1.5,~3,~4.5~)~\upmu$m) do not show any flow vectors inside the pore.
Figure \ref{Figure:3}b.oooo shows evident velocity vector components through initial three pores from left boundary which tends to cease at the fourth pore.
In the magnified view of these pores, Figure \ref{Figure:3}b.oooo($\sigma=~1~\upmu$m) and Figure ($\sigma= 4.5~\upmu$m) have flow magnitude of $1\times10^{-11}$ ms$^{-1}$ towards water phase ( reverse flow) and \ref{Figure:3}b.oooo($\sigma=~2.5~\upmu$m) and ($\sigma= 3.5~\upmu$m) has flow with greater magnitude of $3\times10^{-11}$ ms$^{-1}$ and $2.7\times10^{-11}$ ms$^{-1}$ towards air phase.

\textit{70 nm Multipore Systems:} 
In systems with 70 nm pores, 0FNC state is prominent in the water phase when $\sigma = (2, 4)~\upmu$m. Changing the position of two pores towards the nearest boundaries results in maximum velocity components towards the air phase. A magnified view shows minor flow components inside the pores towards the water phase.
For three-pore systems, there are no flow vectors inside the pores, irrespective of pore position. In four-pore systems, there are no flow components through any of the pores.

The decrease in 0FNC, when the pore size increased from 30 nm to 50 nm is collapsed when the pore size is further increased to 70 nm in Figure \ref{Figure:3}c.
When ($\sigma=~(2,~4)~\upmu$m),
0FNC is prominent in the water phase.
A magnified view of the pores in 
Figure \ref{Figure:3}c.
When ($\sigma=~2~\upmu$m),
and Figure \ref{Figure:3}c.
When ($\sigma=~4~\upmu$m) show no-flow at the interface of the water-air in the pores.
In Figure \ref{Figure:3}c.oo($\sigma=~(1.5,~4)~\upmu$m, the system shows maximum velocity components with magnitude of $3\times10^{-11}$ ms$^{-1}$ through the nanopore towards the air phase.
A magnified view of these pores also shows minor flow components of magnitude $0.3\times10^{-11}$ ms$^{-1}$ inside the pores towards the water phase (reverse flow).
As the number of pores increased to three, there are no flow vectors inside the pores.
In the magnified view of the pores in  Figure \ref{Figure:3}c.ooo($\sigma= (0.3, 0.6, 1.5)~\upmu$m), there are minor flow vectors of magnitude $0.3\times10^{-11}$ ms$^{-1}$ in the water phase and, in Figure \ref{Figure:3}c.ooo($\sigma= (1.5, 3, 4.5 )~\upmu$m) no flow vectors in water phase is observed.
When the number of pores is increased to four, there are no flow components through any of the pores in Figure \ref{Figure:3}c.oooo.
In the magnified view of the pores as in \ref{Figure:3}c.oooo($\sigma= 1.5~\upmu$m)
the velocity vectors are oriented towards the right boundary where as in ($\sigma= 2.5~\upmu$m)
the flow vectors diverges near the nanopore in the water phase.
In Figure \ref{Figure:3}c.oooo($\sigma= 3.5, 4.5~\upmu$m), the flow vectors are oriented to the left boundary in the magnified view of the pore.
The change in orientation of these flow vectors are due to the divergence of flow components as observed in Figure \ref{Figure:3}c.oooo($\sigma= 1, 2.5, 3.5, 4.5~\upmu$m) of velocity streamlines.

The observed behaviour in multi-pore systems differs from single-pore systems, emphasising the complexity of these systems. The influence of pore position in switching between flow and 0FNC state is evident. 
Further investigation is needed to fully understand the dynamics of multi-pore systems.
In four-pore systems with specific pore distances, continuous flow is observed, regardless of pore size for 30 nm and 50 nm pores. However, when pore size is increased to 70 nm and 100 nm, 0FNC states (Appendix Figure 9 and 11) are observed, contrary to single-pore systems. 
This challenges the generalisation from single-pore systems and highlights the importance of pore arrangement.
The complete flow dynamics of multi-pore systems remains complex and requires further investigation.

\subsection*{Analytical approach}
Let us now analytically validate the transport mechanisms and flow structures as shown in Figure \ref{Figure:2}. We used densities of two fluids as $\rho_{g} = 1.225$ kg/m$^{3}$ for air and $\rho_{l} = 997$ kg/m$^{3}$ for water. Since the domain is larger than the wavelength of the instabilities, the velocity potential of the motion is given by Taylor \cite{taylor1950instability} as:

\begin{equation}
    \phi_g = A e^{-k y + nt} \cos kx   
\end{equation}

\begin{equation}
    \phi_l = A e^{k y + nt} \cos kx   
\end{equation}

where $\phi_g$ is the velocity potential in air, $A$ is the amplitude, $k$ is the wave number, $n$ is an integer, and $\phi_l$ is the velocity potential in water. 
We assume that acceleration takes place towards the denser fluid. At the interface,

\begin{equation}
    n^{2} = -k(a_g + a_l) \left(\frac{\rho_{l} - \rho_{g}}{\rho_{l} + \rho_{g}}\right)
\end{equation}

In Equation 5, the terms $a_g$ and $a_l$ represent accelerations in the gaseous and liquid phases, respectively. 
These accelerations can be attributed to the effects of gravity, which are essential in understanding the behaviour of the Rayleigh-Taylor instability. 
The inclusion of these terms is consistent with classical fluid dynamics theory, which accounts for the influence of gravity on fluid flow phenomena.
Moreover, it's important to note that the acceleration terms are critical in determining the growth and behaviour of the instability. 
They represent the rate of change of velocity in response to the density difference between the two phases.
While this explanation doesn't delve into the specifics of how the accelerations are calculated, it establishes their theoretical foundation in the context of Rayleigh-Taylor instability, which is a well-established phenomenon in fluid dynamics.
We acknowledge that in the context of our nanometric geometry, the application of established equations such as the Poiseuille equation and Taylor's characteristic equation may raise questions. 
However, the Poiseuille equation is commonly applied to describe flow in micro- and nanoscale channels under certain conditions, including laminar and fully developed flows \cite{zhou2021wall, sun2021extending, nakad2021taylor, kammara2020development}. 
While our system features nanometric geometries, the use of the Poiseuille equation is justified by the laminar nature of the flow and the steady-state conditions achieved during our simulations. We acknowledge that these conditions are essential for the equation's validity and emphasise that our system meets these criteria.
Taylor's characteristic equation, derived with the presence of an external acceleration superimposed on gravitational acceleration, is applied here to explain the dynamics in our nanoscale system. 
The acceleration terms in the equation account for the effects of gravity on the Rayleigh-Taylor instability, which is essential for our analysis. We acknowledge that the system lacks an external acceleration component but emphasise that the equation's applicability is justified by its utility in elucidating the observed phenomena in our nanoscale setup. 

In our study, we observe the phenomenon of 0FNC state, which represents a unique behaviour within nanofluidic pores. 
This term signifies that despite the absence of conventional macroscopic flow, the system exhibits complex and seemingly chaotic fluid dynamics at the nanoscale.
The use of established equations like the Poiseuille equation and Taylor's characteristic equation may seem unconventional in the context of 0FNC state. 
However, it is important to note that these equations offer valuable insights into the underlying physics of the system, even in situations where macroscopic flow appears to be absent. 
0FNC state is a manifestation of intricate molecular-scale interactions that are not readily captured by conventional continuum mechanics.
While the conditions for applying these equations may seem unusual given the presence of nano-chaotic behaviour, we emphasise that our objective is to gain a deeper understanding of the fundamental principles governing fluid dynamics at the nanoscale. 
The application of established equations serves as a bridge between the macroscopic and molecular realms, allowing us to uncover hidden complexities in nanofluidic systems.
A single nanofluidic pore with diameter $d = 30$ nm and maximum flow velocity will have $n^{2} = -1.2\times10^{-6}k$. 
This notation was chosen for its convenience in representing the relationship between these variables in our context. 
While unconventional, it accurately reflects the key aspects of our analysis and allows for a more intuitive understanding of the system.
Since $(a_g + a_l)$ is positive, The flow instability evolves over time, initially exhibiting transient behaviour until it stabilises into a consistent pattern. This behaviours can be described by a relationship proportional to $\sqrt{\frac{\rho_l - \rho_g}{\rho_l+ \rho_g}}$, which is unity in our system and explains the no-flow systems. For systems without flow, Equation (2) simplifies to:

\begin{equation}
    n^{2} = 0
\end{equation}

In the water phase, the displacement of instability is given by:

\begin{equation}
    s = \frac{1}{2} a_g t^{2}
\end{equation}

where $t$ is the time taken for the steady state to be reached from the initial time, $a_g$ is the acceleration of the molecules in the gaseous phase, and $\bm v$ is the velocity of flow. In an evaporating liquid-gas interface, a certain number of molecules condense onto the surface of the liquid, which can result in backflow in some cases. The ratio of atoms condensing to atoms evaporating is unclear and can be complicated to predict from system to system. The number of condensing molecules is a function of the mean velocity through the nanofluidic pore. The number of evaporating particles can be described by Feynman \cite{feynman1965feynman} as:

\begin{equation}
    N = \frac{1}{\Gamma} \frac{\nu}{D}e^{-W/k_B T}
\end{equation}

where $\nu$ is the average speed of particles, $W$ is the extra energy needed for evaporation, $k_B$ is the Boltzmann constant, and $\Gamma$ is the cross-sectional area. However, since we describe the system using continuum mechanics, we do not pursue this quantum mechanical perspective.
The vapour transport through the nanofluidic pore is governed by the conventional Hagen-Poiseuille equation \cite{lu2015modeling}:

\begin{equation}
    \Delta {P} = P_{in} - P_{l} = \frac{8 \eta L \dot{m} }{\pi r^4 \rho_l}
\end{equation}

where $P_{i}$ is the pressure at the interface, $P_{l}$ is the liquid phase pressure, $\dot{m}$ is the mass flow rate through the nanofluidic pore, $\eta$ is the viscosity of the liquid, $r$ is the radius of the nanofluidic pore, $L$ is the flow length, and $\rho$ is the liquid density. Since we are addressing transport in nanofluidic pores using the Rayleigh-Taylor instability:

\begin{equation}
     \Delta {P_s} = \frac{4\eta {g_g t^2 \dot{m}}}{\pi r^4 \rho_l} = \frac{8 \eta \alpha_e N_{el} \dot{m}}{\pi r^4 \rho_l}
\end{equation}

where $\Delta {P_s}$ is the pressure drop in simulations, $\alpha_{e}$ is the element size of the geometry used for simulation, and $N_{e}$ is the number of elements in the flow region. Then the simulation velocity is given by:

\begin{equation}
    \bm{v}_{s} = \frac{\Delta{P_s} \pi r^4 \rho_l t}{8 \eta \dot{m}}
\end{equation}

The analytical form of $\Delta {P}$ is given by:

\begin{equation}
    \Delta {P_a} = \frac{4 \eta \bm{v}_a \sigma \Phi}{\pi r^4}
\end{equation}

and the corresponding velocity is given by:

\begin{equation}
    \bm{v}_{a} = \frac{\Delta {P_a} \pi r^4 }{4 {\eta} \sigma \Phi}
\end{equation}

where ${\sigma}$ is the distance of the nanopore from the boundary of the system, $\Phi$ is the area of the nanopore, $\Delta{P_a}$ is the pressure drop across the interface, which is a constant. In Figure \ref{Figure:4}g-i, we observe certain agreements between analytical and numerical velocities, except in a few cases where large errors are observed.


\section*{Discussion}
\begin{figure}[]
\includegraphics[width =0.45\textwidth]{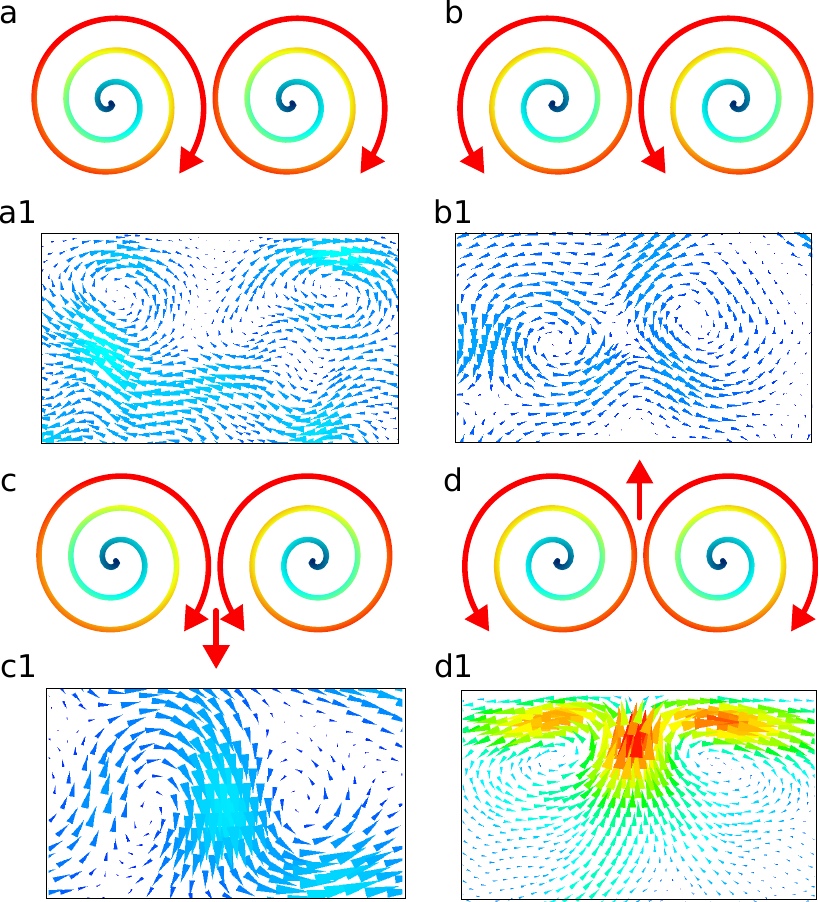}
\caption{\textbf{Geometry dependent pairwise-vortices driven flow.} (a) Schematic of a pair of clockwise votices with zero resultant flow vector.
(a1) Results from the simulation with pair unidirectrional clockwise votices without any resultant flow vector.
(b) Schematic of pair of anticlockwise vortices with zero resultant flow vector. 
(b1) Results from the simulation with pair unidirectrional anticlockwise votices without any resultant flow vector.
(c) Schematic of resultant flow vector for pair of clockwise and anticlockwise vortices.
(c1) Simulation following the prediction in (c).
(d) Schematic of strong resultant flow vector for pair of anticlockwise and clockwise vortices.
(d1) Simulation showing strong resultant flow vector as depicted in (d).
} 
\label{Figure:5}
\end{figure}

In this nanofluidic study of Rayleigh-Taylor instability, we found two unique phenomena -- self-driven flow and 0FNC state. 
In Figure \ref{Figure:3}, some velocity vectors show flow in both directions of water and air without applying any external driving force with a high probability of flow in the direction against the gravity from water to air.
Figure \ref{Figure:4}a.o(d= 70 nm, $\sigma= 1.7~\upmu$m) represents the systems where the cloud of high-velocity molecules exist and the velocity vectors are pointed towards the air volume fraction.
Figure \ref{Figure:4}b.o(d= 70 nm, $\sigma= 2.5~\upmu$m) represents the systems with no flow in the air and the strong back-flow in water creates vertices throughout the boundaries in the water phase.
The velocity vectors show several asymmetric swirls and sudden changes in directions with a turbulent nature. 
In Figure \ref{Figure:4}c.o(d= 50 nm, $\sigma= 2.8~\upmu$m) represent a system with a significant flow difference pattern in the air phase.
There are also cases where back-flow towards the water phase is initiated but is relatively less prominent than the velocity vectors towards the water phase.
The velocity stream lines in Figure \ref{Figure:4}d.o(d= 70 nm, $\sigma= 1.7~\upmu$m) shows highest magnitude of $8.66\times10^{-12}$ ms$^{-1}$ at the nanopore in the water phase.
When the position of nanopore changes to $\sigma= 2.5~\upmu$m, in Figure \ref{Figure:4}e.o(d= 70 nm, $\sigma= 2.5~\upmu$m) the highest  magnitude of velocity streamlines is $4.35\times10^{-12}$ ms$^{-1}$.
In Figure \ref{Figure:4}f.o(d= 50 nm, $\sigma= 2.8~\upmu$m), the velocity streamlines are prominent through the nanopore with a magnitude of $8.66\times10^{-12}$ ms$^{-1}$.
In Figure \ref{Figure:3}, the velocity distribution of the multiple nanofluidic pore system become more complex than the single nanofluidic pore system.
This is indicative that the molecular species in the (NVT) ensemble distribute their kinetic energy in a way to maintain the number of evaporated molecules and the number of condensing molecules in equilibrium.
The vortex pairs and their vector orientation in these systems are very significant in determining net flow through the nanopores.
The disruption of equilibrium at the interface and the backflow through the nanofluidic pore to the water domain may lead to the phenomenon of cavitation \cite{sperry2003evolution} which hinders the further flow which is not uncommon in transport through confined nanofluidic pore systems.
Usually, the variations in contact angle and the liquid pressure at the interface control the evaporation process to a great extent and heat is to be supplied for it to sustain \cite{lu2015modeling}.
Figure \ref{Figure:5} gives the schematic representation of self-driven flow in these system,
In Figure \ref{Figure:5}a and Figure \ref{Figure:5}b, there is no net flow due to the cancellation of equal and opposite vector components.
Whereas in Figure \ref{Figure:5}c and \ref{Figure:5}d, the resultant position vector is given by,
\begin{equation}
    \bm{r} = {b}{\theta} + {b_1}{\theta_1}
\end{equation}
where $b$ and ${b_1}$ determines the distance between the spiral loops of each spiral within the vortex pair spirals.
Then Equation 13 with the correction term becomes,
\begin{equation}
      \bm{v}_{a} = \frac{\Delta {P_a} \pi \bm{r}^4 }{4 {\eta} \sigma \Phi} + \sqrt{{\bm{\beta}^2} + {\omega^2}|{b}{\theta} + {b_1}{\theta_1}|^2}
\end{equation}
and for systems with no flow, Equation 13 becomes,
\begin{equation}
       \bm{v}_{a} = \frac{\Delta {P_a} \pi r^4 }{4 {\eta} \sigma \Phi} + |\bm{\beta}|
\end{equation}
where $\bm{\beta}$ is the  constant velocity component along the perpendicular direction of nanopore throat.
These vortex pairs are evident throughout every simulated system regardless of $d$ and $\sigma$ (as in Figure 5. a1 - d1).

We recognise the potential concern regarding the validity of adopting a continuum approach in a system with a large Knudsen number. 
The Knudsen number, which characterises the ratio of molecular mean free path to characteristic length scale, is indeed an important parameter in micro and nanoscale fluid dynamics. 
In our study, we have employed a continuum approach as it provides valuable insights into the behaviour of nanofluidic flows under the conditions we have considered. 
While it is true that the Knudsen number for our system can be large, and slip or rarefied gas effects may become significant in such cases, it is important to note that the continuum approach serves as a valuable first step in understanding the underlying physics.
We also recognise that the slip regime and rarefied gas effects are intriguing topics in nanofluidic research, and we plan to explore these aspects in future investigations. 
The current study lays the foundation for understanding the complex fluid dynamics at the nanoscale, and our findings contribute to the broader understanding of nanofluidic phenomena. 
The significant reduction in surface tension and surface conductivity in nanopores (compared to that of bulk liquid) is a  frequent topic of discussion {\cite{knight2019water, takei2000changes}}.
The decreasing pore size to nanometre length-scale impacts the hydrogen bonding network in water as a result has lower coordination number of bonding water molecule than the bulk water \cite{senanayake2021simulations}.
Due to this enhanced affinity of water molecules for the silica surface has a large impact on the system's molecular dynamics and makes the relationship between the effects of surface tension and nanoconfinement much smaller {\cite{galteland2019pressures}}.
In the modified Laplace-Young equation as in {\cite{qiao2009pressurized}} which considers the solid-liquid interfacial tension, thermal energy exchange, and variation in molecular dynamics in a nanofluidic channel, the pressure gradient of whole system has significant dependency on diameter of transport channel, and the pressure is in the range of megapascal.
This changes when phase transition occurs and the nanochannel is connected to a bulk liquid, which has dimensions much greater to that of transport channel as in our case (maximum width of channel $=$ 70 nm, maximum length of channel $=$ 100 nm, maximum length of bulk reservoir $=$ 6 $\upmu$m).
Magda et al. shows in Lennard–Jones liquid of  2-12 moleculs-wide pores that the solvation force on interfacial pores of 9.5 liquid diameter wide pore are equal to the pressure of bulk liquid which is in equilibrium with the pore \cite{magda1985molecular}.
However, this may vary for interfacial tension of pores with smaller width less than 9.5 liquid diameter depending on the pressure {\cite{magda1985molecular}.
Similarly, the effect of breakdown of Navier-Stokes formulations and equation of continuity on flow dynamics of microfluidic systems needs to be considered, which is addressed in a following study by us {\cite{johny2022superfluidic}}.
However, further study on time-dependent transient states needs to be explored for obtaining greater insight of the exact time-frames and boundary condition to determine the exact point when continuum breaks, which is highly computationally expensive and requires a separate attention. 

In this study, we consider the mean free path and Knudsen number as a prerequisite to the physics of the upcoming studies.
We have built a basic structure of a nanoconfined system with critical values to address the physics of molecular nanofluids which is at the classical-quantum interface.
Hence, at this point it may not be appropriate to generalise the results with a critical parameter or non-dimensional constant.
The main objective of this study is to use computational and theoretical means to not only predict the molecular fluid dynamics but also create a potentially innovative tool of solid-state nanopore's fluid dynamics. 
While our work primarily employs continuum mechanics and analytical methods, it is noteworthy that similar phenomena have been observed using MD simulations in prior research \cite{zou2020transpiration}. 
Due to the significant computational complexities associated with MD simulations at the scales considered in our study, we opted for a continuum approach.  
Our results follow the finding of Zou et al. \cite{zou2020transpiration} and extends it beyond the scale and capacity of their study while finding global impact of nanopores in a bulk system. 
We use the term `zero-flow nano-chaos' to describe a unique phenomenon observed in our nanoscale fluid dynamics, which exhibits complex and seemingly chaotic behaviour despite the absence of classical turbulence. 
It is important to note that this terminology is used to characterise the specific flow patterns and instabilities observed in our system and does not imply that we are modelling or simulating classical turbulence, as our adopted model does not explicitly account for turbulence effects.


In short, this study of nanofluidic systems unveils two intriguing phenomena: self-driven flow and 0FNC state. Self-driven flow, as observed in Figure \ref{Figure:3}, reveals the unexpected occurrence of fluid motion without the need for external driving forces. Notably, we found a high probability of flow against gravity, from water to air, challenging conventional expectations.
Figure \ref{Figure:4}a.o demonstrates the presence of high-velocity molecular clouds, where velocity vectors point towards the air phase. Conversely, Figure \ref{Figure:4}b.o illustrates situations where air remains stagnant, resulting in strong backflow within the water phase, creating vortex structures along the boundaries. These velocity vectors exhibit asymmetrical swirls and sudden directional changes, indicative of turbulent behaviour.
Figure \ref{Figure:4}c.o presents a system with significant airflow differences in the air phase. While instances of backflow into the water phase do occur, they are less prominent than the vectors directed towards the water phase. Notably, the highest recorded velocity magnitude of $8.66\times10^{-12}$ ms$^{-1}$ at the nanopore occurs in Figure \ref{Figure:4}d.o.
The complex nature of the velocity distribution in systems with multiple nanofluidic pores, as shown in Figure \ref{Figure:3}, underscores the role of molecular species' kinetic energy distribution in maintaining equilibrium between evaporated and condensing molecules. The formation and orientation of vortex pairs are pivotal in determining the net flow through these nanopores. 
The potential for disruption of equilibrium at the interface and backflow through the nanofluidic pore into the water domain raises intriguing possibilities, such as cavitation \cite{sperry2003evolution}, which can hinder further flow—a phenomenon not uncommon in confined nanofluidic pore systems.
Our study sheds light on the intricate interplay between surface tension, conductivity, and confinement in nanofluidic systems. The reduction in surface tension and conductivity at the nanoscale \cite{knight2019water, takei2000changes} significantly impacts the hydrogen bonding network in water, resulting in a lower coordination number of bonded water molecules compared to bulk water \cite{senanayake2021simulations}. This enhanced affinity of water for silica surfaces influences molecular dynamics, with important implications for pressure gradients \cite{qiao2009pressurized}.
While our research contributes to understanding fluid dynamics at the classical-quantum interface, it's essential to recognise that our findings are preliminary. The complex nature of these systems, encompassing components that defy conventional viscosity, hints at the interplay between anharmonic and harmonic elements within the same system at room temperature \cite{johny2023quantum}. 
To conclude, our study explores the non-trivial behaviour of Rayleigh-Taylor instability in sub-100 nanometer fluidic pores, showcasing the potential for self-driven flow and flow control without external forces. These discoveries open doors to fundamental research in nanoscale fluid dynamics and bear relevance to fields such as sustainable energy production, nanorobotics, biomolecular diagnostics, and the broader physics of fluid dynamics.

\section*{Acknowledgements}
The authors are grateful to the internal funding of the International Center for Nanodevices. 
Authors are thankful to the funding provided by Honeywell Corporate Social Responsibility grant; VJ was funded by this funding.
SG thanks the German Research Foundation for partly supporting this research.
Professor Natalia Berloff's scientific and mathematical opinion have found several important findings -- sincere acknowledgement for this intellectual contribution in this work.
The authors are thankful to Dr Moumita Ghosh for many productive discussion. 
Jintu K James have validated the simulations independently; we are thankful to him for that.





\bibliography{rsc} 

\providecommand*{\mcitethebibliography}{\thebibliography}
\csname @ifundefined\endcsname{endmcitethebibliography}
{\let\endmcitethebibliography\endthebibliography}{}
\begin{mcitethebibliography}{39}
\providecommand*{\natexlab}[1]{#1}
\providecommand*{\mciteSetBstSublistMode}[1]{}
\providecommand*{\mciteSetBstMaxWidthForm}[2]{}
\providecommand*{\mciteBstWouldAddEndPuncttrue}
  {\def\EndOfBibitem{\unskip.}}
\providecommand*{\mciteBstWouldAddEndPunctfalse}
  {\let\EndOfBibitem\relax}
\providecommand*{\mciteSetBstMidEndSepPunct}[3]{}
\providecommand*{\mciteSetBstSublistLabelBeginEnd}[3]{}
\providecommand*{\EndOfBibitem}{}
\mciteSetBstSublistMode{f}
\mciteSetBstMaxWidthForm{subitem}
{(\emph{\alph{mcitesubitemcount}})}
\mciteSetBstSublistLabelBeginEnd{\mcitemaxwidthsubitemform\space}
{\relax}{\relax}

\bibitem[Ali \emph{et~al.}(2017)Ali, Ahmed, Ramirez, Nasir, Cervera, Mafe,
  Niemeyer, and Ensinger]{ali2017cesium}
M.~Ali, I.~Ahmed, P.~Ramirez, S.~Nasir, J.~Cervera, S.~Mafe, C.~M. Niemeyer and
  W.~Ensinger, \emph{Langmuir}, 2017, \textbf{33}, 9170--9177\relax
\mciteBstWouldAddEndPuncttrue
\mciteSetBstMidEndSepPunct{\mcitedefaultmidpunct}
{\mcitedefaultendpunct}{\mcitedefaultseppunct}\relax
\EndOfBibitem
\bibitem[Zhang \emph{et~al.}(2020)Zhang, Huang, Qian, Chen, Wen, and
  Jiang]{zhang2020engineering}
Z.~Zhang, X.~Huang, Y.~Qian, W.~Chen, L.~Wen and L.~Jiang, \emph{Advanced
  Materials}, 2020, \textbf{32}, 1904351\relax
\mciteBstWouldAddEndPuncttrue
\mciteSetBstMidEndSepPunct{\mcitedefaultmidpunct}
{\mcitedefaultendpunct}{\mcitedefaultseppunct}\relax
\EndOfBibitem
\bibitem[Ghosh \emph{et~al.}(2020)Ghosh, Karedla, and Gregor]{ghosh2020single}
S.~Ghosh, N.~Karedla and I.~Gregor, \emph{Lab on a Chip}, 2020, \textbf{20},
  3249--3257\relax
\mciteBstWouldAddEndPuncttrue
\mciteSetBstMidEndSepPunct{\mcitedefaultmidpunct}
{\mcitedefaultendpunct}{\mcitedefaultseppunct}\relax
\EndOfBibitem
\bibitem[Zou \emph{et~al.}(2020)Zou, Gupta, and Maroo]{zou2020transpiration}
A.~Zou, M.~Gupta and S.~C. Maroo, \emph{The journal of physical chemistry
  letters}, 2020, \textbf{11}, 3637--3641\relax
\mciteBstWouldAddEndPuncttrue
\mciteSetBstMidEndSepPunct{\mcitedefaultmidpunct}
{\mcitedefaultendpunct}{\mcitedefaultseppunct}\relax
\EndOfBibitem
\bibitem[Siria \emph{et~al.}(2013)Siria, Poncharal, Biance, Fulcrand, Blase,
  Purcell, and Bocquet]{siria2013giant}
A.~Siria, P.~Poncharal, A.-L. Biance, R.~Fulcrand, X.~Blase, S.~T. Purcell and
  L.~Bocquet, \emph{Nature}, 2013, \textbf{494}, 455--458\relax
\mciteBstWouldAddEndPuncttrue
\mciteSetBstMidEndSepPunct{\mcitedefaultmidpunct}
{\mcitedefaultendpunct}{\mcitedefaultseppunct}\relax
\EndOfBibitem
\bibitem[Siria \emph{et~al.}(2017)Siria, Bocquet, and Bocquet]{Bocquet2017}
A.~Siria, M.-L. Bocquet and L.~Bocquet, \emph{Nature Reviews Chemistry}, 2017,
  \textbf{1}, 0091\relax
\mciteBstWouldAddEndPuncttrue
\mciteSetBstMidEndSepPunct{\mcitedefaultmidpunct}
{\mcitedefaultendpunct}{\mcitedefaultseppunct}\relax
\EndOfBibitem
\bibitem[Heydt and Steudle(1991)]{heydt1991measurement}
H.~Heydt and E.~Steudle, \emph{Planta}, 1991, \textbf{184}, 389--396\relax
\mciteBstWouldAddEndPuncttrue
\mciteSetBstMidEndSepPunct{\mcitedefaultmidpunct}
{\mcitedefaultendpunct}{\mcitedefaultseppunct}\relax
\EndOfBibitem
\bibitem[Gates(1968)]{gates1968transpiration}
D.~M. Gates, \emph{Annual Review of Plant Physiology}, 1968, \textbf{19},
  211--238\relax
\mciteBstWouldAddEndPuncttrue
\mciteSetBstMidEndSepPunct{\mcitedefaultmidpunct}
{\mcitedefaultendpunct}{\mcitedefaultseppunct}\relax
\EndOfBibitem
\bibitem[Rashidi \emph{et~al.}(2014)Rashidi, Freidoonimehr, Hosseini, B{\'e}g,
  and Hung]{rashidi2014homotopy}
M.~Rashidi, N.~Freidoonimehr, A.~Hosseini, O.~A. B{\'e}g and T.-K. Hung,
  \emph{Meccanica}, 2014, \textbf{49}, 469--482\relax
\mciteBstWouldAddEndPuncttrue
\mciteSetBstMidEndSepPunct{\mcitedefaultmidpunct}
{\mcitedefaultendpunct}{\mcitedefaultseppunct}\relax
\EndOfBibitem
\bibitem[Skaug \emph{et~al.}(2018)Skaug, Schwemmer, Fringes, Rawlings, and
  Knoll]{skaug2018nanofluidic}
M.~J. Skaug, C.~Schwemmer, S.~Fringes, C.~D. Rawlings and A.~W. Knoll,
  \emph{Science}, 2018, \textbf{359}, 1505--1508\relax
\mciteBstWouldAddEndPuncttrue
\mciteSetBstMidEndSepPunct{\mcitedefaultmidpunct}
{\mcitedefaultendpunct}{\mcitedefaultseppunct}\relax
\EndOfBibitem
\bibitem[Ahmed \emph{et~al.}(2020)Ahmed, Ramesan, Lee, Rezk, and
  Yeo]{ahmed2020chip}
H.~Ahmed, S.~Ramesan, L.~Lee, A.~R. Rezk and L.~Y. Yeo, \emph{Small}, 2020,
  \textbf{16}, 1903605\relax
\mciteBstWouldAddEndPuncttrue
\mciteSetBstMidEndSepPunct{\mcitedefaultmidpunct}
{\mcitedefaultendpunct}{\mcitedefaultseppunct}\relax
\EndOfBibitem
\bibitem[Ruggeri \emph{et~al.}(2017)Ruggeri, Zosel, Mutter, R{\'o}{\.z}ycka,
  Wojtas, O{\.z}yhar, Schuler, and Krishnan]{ruggeri2017single}
F.~Ruggeri, F.~Zosel, N.~Mutter, M.~R{\'o}{\.z}ycka, M.~Wojtas, A.~O{\.z}yhar,
  B.~Schuler and M.~Krishnan, \emph{Nature Nanotechnology}, 2017, \textbf{12},
  488--495\relax
\mciteBstWouldAddEndPuncttrue
\mciteSetBstMidEndSepPunct{\mcitedefaultmidpunct}
{\mcitedefaultendpunct}{\mcitedefaultseppunct}\relax
\EndOfBibitem
\bibitem[Cira \emph{et~al.}(2015)Cira, Benusiglio, and Prakash]{cira2015vapour}
N.~J. Cira, A.~Benusiglio and M.~Prakash, \emph{Nature}, 2015, \textbf{519},
  446--450\relax
\mciteBstWouldAddEndPuncttrue
\mciteSetBstMidEndSepPunct{\mcitedefaultmidpunct}
{\mcitedefaultendpunct}{\mcitedefaultseppunct}\relax
\EndOfBibitem
\bibitem[Cheng \emph{et~al.}(1997)Cheng, Van~Hoek, Yeager, Verkman, and
  Mitra]{cheng1997three}
A.~Cheng, A.~Van~Hoek, M.~Yeager, A.~Verkman and A.~Mitra, \emph{Nature}, 1997,
  \textbf{387}, 627--630\relax
\mciteBstWouldAddEndPuncttrue
\mciteSetBstMidEndSepPunct{\mcitedefaultmidpunct}
{\mcitedefaultendpunct}{\mcitedefaultseppunct}\relax
\EndOfBibitem
\bibitem[de~Groot and Grubmuller(2001)]{de2001water}
B.~L. de~Groot and H.~Grubmuller, \emph{Science}, 2001, \textbf{294},
  2353--2357\relax
\mciteBstWouldAddEndPuncttrue
\mciteSetBstMidEndSepPunct{\mcitedefaultmidpunct}
{\mcitedefaultendpunct}{\mcitedefaultseppunct}\relax
\EndOfBibitem
\bibitem[Wang \emph{et~al.}(2011)Wang, Cui, Sun, and Zhang]{wang2011tunneling}
Y.~Wang, J.~Cui, X.~Sun and Y.~Zhang, \emph{Cell Death \& Differentiation},
  2011, \textbf{18}, 732--742\relax
\mciteBstWouldAddEndPuncttrue
\mciteSetBstMidEndSepPunct{\mcitedefaultmidpunct}
{\mcitedefaultendpunct}{\mcitedefaultseppunct}\relax
\EndOfBibitem
\bibitem[Sharp(1984)]{sharp1984overview}
D.~H. Sharp, \emph{Physica D: Nonlinear Phenomena}, 1984, \textbf{12},
  3--18\relax
\mciteBstWouldAddEndPuncttrue
\mciteSetBstMidEndSepPunct{\mcitedefaultmidpunct}
{\mcitedefaultendpunct}{\mcitedefaultseppunct}\relax
\EndOfBibitem
\bibitem[Linden \emph{et~al.}(1994)Linden, Redondo, and
  Youngs]{linden1994molecular}
P.~Linden, J.~Redondo and D.~Youngs, \emph{Journal of Fluid Mechanics}, 1994,
  \textbf{265}, 97--124\relax
\mciteBstWouldAddEndPuncttrue
\mciteSetBstMidEndSepPunct{\mcitedefaultmidpunct}
{\mcitedefaultendpunct}{\mcitedefaultseppunct}\relax
\EndOfBibitem
\bibitem[Taylor(1950)]{taylor1950instability}
G.~I. Taylor, \emph{Proceedings of the Royal Society of London. Series A.
  Mathematical and Physical Sciences}, 1950, \textbf{201}, 192--196\relax
\mciteBstWouldAddEndPuncttrue
\mciteSetBstMidEndSepPunct{\mcitedefaultmidpunct}
{\mcitedefaultendpunct}{\mcitedefaultseppunct}\relax
\EndOfBibitem
\bibitem[Meller and Branton(2002)]{meller2002single}
A.~Meller and D.~Branton, \emph{Electrophoresis}, 2002, \textbf{23},
  2583--2591\relax
\mciteBstWouldAddEndPuncttrue
\mciteSetBstMidEndSepPunct{\mcitedefaultmidpunct}
{\mcitedefaultendpunct}{\mcitedefaultseppunct}\relax
\EndOfBibitem
\bibitem[Xuan and Li(2007)]{xuan2007solute}
X.~Xuan and D.~Li, \emph{Electrophoresis}, 2007, \textbf{28}, 627--634\relax
\mciteBstWouldAddEndPuncttrue
\mciteSetBstMidEndSepPunct{\mcitedefaultmidpunct}
{\mcitedefaultendpunct}{\mcitedefaultseppunct}\relax
\EndOfBibitem
\bibitem[Yoder \emph{et~al.}(1994)Yoder, Ryan, Waring, Schoettle, and
  Kaufmann]{10.1093/forestscience/40.3.513}
B.~J. Yoder, M.~G. Ryan, R.~H. Waring, A.~W. Schoettle and M.~R. Kaufmann,
  \emph{Forest Science}, 1994, \textbf{40}, 513--527\relax
\mciteBstWouldAddEndPuncttrue
\mciteSetBstMidEndSepPunct{\mcitedefaultmidpunct}
{\mcitedefaultendpunct}{\mcitedefaultseppunct}\relax
\EndOfBibitem
\bibitem[Narayanan \emph{et~al.}(2011)Narayanan, Fedorov, and
  Joshi]{narayanan2011interfacial}
S.~Narayanan, A.~G. Fedorov and Y.~K. Joshi, \emph{Langmuir}, 2011,
  \textbf{27}, 10666--10676\relax
\mciteBstWouldAddEndPuncttrue
\mciteSetBstMidEndSepPunct{\mcitedefaultmidpunct}
{\mcitedefaultendpunct}{\mcitedefaultseppunct}\relax
\EndOfBibitem
\bibitem[Zhou \emph{et~al.}(2021)Zhou, Sun, and Bai]{zhou2021wall}
R.~Zhou, C.~Sun and B.~Bai, \emph{The Journal of Chemical Physics}, 2021,
  \textbf{154}, 1--2\relax
\mciteBstWouldAddEndPuncttrue
\mciteSetBstMidEndSepPunct{\mcitedefaultmidpunct}
{\mcitedefaultendpunct}{\mcitedefaultseppunct}\relax
\EndOfBibitem
\bibitem[Sun \emph{et~al.}(2021)Sun, Zhou, Zhao, and Bai]{sun2021extending}
C.~Sun, R.~Zhou, Z.~Zhao and B.~Bai, \emph{Langmuir}, 2021, \textbf{37},
  6158--6167\relax
\mciteBstWouldAddEndPuncttrue
\mciteSetBstMidEndSepPunct{\mcitedefaultmidpunct}
{\mcitedefaultendpunct}{\mcitedefaultseppunct}\relax
\EndOfBibitem
\bibitem[Nakad \emph{et~al.}(2021)Nakad, Witelski, Domec, Sevanto, and
  Katul]{nakad2021taylor}
M.~Nakad, T.~Witelski, J.-C. Domec, S.~Sevanto and G.~Katul, \emph{Journal of
  Fluid Mechanics}, 2021, \textbf{913}, A44\relax
\mciteBstWouldAddEndPuncttrue
\mciteSetBstMidEndSepPunct{\mcitedefaultmidpunct}
{\mcitedefaultendpunct}{\mcitedefaultseppunct}\relax
\EndOfBibitem
\bibitem[Kammara and Kumar(2020)]{kammara2020development}
K.~K. Kammara and R.~Kumar, \emph{Microfluidics and Nanofluidics}, 2020,
  \textbf{24}, 70\relax
\mciteBstWouldAddEndPuncttrue
\mciteSetBstMidEndSepPunct{\mcitedefaultmidpunct}
{\mcitedefaultendpunct}{\mcitedefaultseppunct}\relax
\EndOfBibitem
\bibitem[Feynman \emph{et~al.}(1965)Feynman, Leighton, and
  Sands]{feynman1965feynman}
R.~P. Feynman, R.~B. Leighton and M.~Sands, \emph{American Journal of Physics},
  1965, \textbf{33}, 750--752\relax
\mciteBstWouldAddEndPuncttrue
\mciteSetBstMidEndSepPunct{\mcitedefaultmidpunct}
{\mcitedefaultendpunct}{\mcitedefaultseppunct}\relax
\EndOfBibitem
\bibitem[Lu \emph{et~al.}(2015)Lu, Narayanan, and Wang]{lu2015modeling}
Z.~Lu, S.~Narayanan and E.~N. Wang, \emph{Langmuir}, 2015, \textbf{31},
  9817--9824\relax
\mciteBstWouldAddEndPuncttrue
\mciteSetBstMidEndSepPunct{\mcitedefaultmidpunct}
{\mcitedefaultendpunct}{\mcitedefaultseppunct}\relax
\EndOfBibitem
\bibitem[Sperry(2003)]{sperry2003evolution}
J.~S. Sperry, \emph{International Journal of Plant Sciences}, 2003,
  \textbf{164}, S115--S127\relax
\mciteBstWouldAddEndPuncttrue
\mciteSetBstMidEndSepPunct{\mcitedefaultmidpunct}
{\mcitedefaultendpunct}{\mcitedefaultseppunct}\relax
\EndOfBibitem
\bibitem[Knight \emph{et~al.}(2019)Knight, Kalugin, Coker, and
  Ilgen]{knight2019water}
A.~W. Knight, N.~G. Kalugin, E.~Coker and A.~G. Ilgen, \emph{Scientific
  reports}, 2019, \textbf{9}, 8246\relax
\mciteBstWouldAddEndPuncttrue
\mciteSetBstMidEndSepPunct{\mcitedefaultmidpunct}
{\mcitedefaultendpunct}{\mcitedefaultseppunct}\relax
\EndOfBibitem
\bibitem[Takei \emph{et~al.}(2000)Takei, Mukasa, Kofuji, Fuji, Watanabe,
  Chikazawa, and Kanazawa]{takei2000changes}
T.~Takei, K.~Mukasa, M.~Kofuji, M.~Fuji, T.~Watanabe, M.~Chikazawa and
  T.~Kanazawa, \emph{Colloid and Polymer Science}, 2000, \textbf{278},
  475--480\relax
\mciteBstWouldAddEndPuncttrue
\mciteSetBstMidEndSepPunct{\mcitedefaultmidpunct}
{\mcitedefaultendpunct}{\mcitedefaultseppunct}\relax
\EndOfBibitem
\bibitem[Senanayake \emph{et~al.}(2021)Senanayake, Greathouse, Ilgen, and
  Thompson]{senanayake2021simulations}
H.~S. Senanayake, J.~A. Greathouse, A.~G. Ilgen and W.~H. Thompson, \emph{The
  Journal of Chemical Physics}, 2021, \textbf{154}, 104503\relax
\mciteBstWouldAddEndPuncttrue
\mciteSetBstMidEndSepPunct{\mcitedefaultmidpunct}
{\mcitedefaultendpunct}{\mcitedefaultseppunct}\relax
\EndOfBibitem
\bibitem[Galteland \emph{et~al.}(2019)Galteland, Bedeaux, Hafskjold, and
  Kjelstrup]{galteland2019pressures}
O.~Galteland, D.~Bedeaux, B.~Hafskjold and S.~Kjelstrup, \emph{Frontiers in
  Physics}, 2019, \textbf{7}, 60\relax
\mciteBstWouldAddEndPuncttrue
\mciteSetBstMidEndSepPunct{\mcitedefaultmidpunct}
{\mcitedefaultendpunct}{\mcitedefaultseppunct}\relax
\EndOfBibitem
\bibitem[Qiao \emph{et~al.}(2009)Qiao, Liu, and Chen]{qiao2009pressurized}
Y.~Qiao, L.~Liu and X.~Chen, \emph{Nano letters}, 2009, \textbf{9},
  984--988\relax
\mciteBstWouldAddEndPuncttrue
\mciteSetBstMidEndSepPunct{\mcitedefaultmidpunct}
{\mcitedefaultendpunct}{\mcitedefaultseppunct}\relax
\EndOfBibitem
\bibitem[Magda \emph{et~al.}(1985)Magda, Tirrell, and
  Davis]{magda1985molecular}
J.~Magda, M.~Tirrell and H.~Davis, \emph{The Journal of chemical physics},
  1985, \textbf{83}, 1888--1901\relax
\mciteBstWouldAddEndPuncttrue
\mciteSetBstMidEndSepPunct{\mcitedefaultmidpunct}
{\mcitedefaultendpunct}{\mcitedefaultseppunct}\relax
\EndOfBibitem
\bibitem[Johny \emph{et~al.}(2022)Johny, Contera, and
  Ghosh]{johny2022superfluidic}
V.~Johny, S.~A. Contera and S.~Ghosh, \emph{arXiv preprint arXiv:2208.13759},
  2022\relax
\mciteBstWouldAddEndPuncttrue
\mciteSetBstMidEndSepPunct{\mcitedefaultmidpunct}
{\mcitedefaultendpunct}{\mcitedefaultseppunct}\relax
\EndOfBibitem
\bibitem[Johny and Ghosh(2023)]{johny2023quantum}
V.~Johny and S.~Ghosh, \emph{ChemRxiv}, 2023\relax
\mciteBstWouldAddEndPuncttrue
\mciteSetBstMidEndSepPunct{\mcitedefaultmidpunct}
{\mcitedefaultendpunct}{\mcitedefaultseppunct}\relax
\EndOfBibitem
\bibitem[Manual(2009)]{manual2009ansys}
U.~Manual, \emph{Theory Guide}, 2009\relax
\mciteBstWouldAddEndPuncttrue
\mciteSetBstMidEndSepPunct{\mcitedefaultmidpunct}
{\mcitedefaultendpunct}{\mcitedefaultseppunct}\relax
\EndOfBibitem
\end{mcitethebibliography}
\bibliographystyle{rsc} 

\clearpage

\newpage

\appendix
\section{Notations}
\begin{table}[h!]
\begin{tabular}{ l  l }
\hline
 $\sigma$& Position of nanopore from the reference edge \\
 $d$& Diameter of the Nanopore
\\
 $\gamma$& Phase volume fraction in the multiphase flow\\
 $\rho$& Phase density\\
 $\kappa$& Phase variable\\
 $\bm v$& Phase velocity\\
 $M$& Mass transport through the nanopore\\
 $\rttensor\tau$& Diffusion term/stress tensor\\
 $\dot{m}_{lg}$& Mass transfer rate from phase $l$ to phase $g$\\
 $\dot{m}_{gl}$& Mass transfer rate from phase $g$ to phase $l$\\
 $\phi$& Velocity potential\\
 $A$& Amplitude of the wave\\
 $k$& Wave number\\
 $a$& Acceleration of the molecules\\
 $N$& Number of evaporating particles from the nanopore\\
 $W$& Energy corresponding to work function of evaporation\\
 $k_B$& Boltzmann constant\\
 $\Gamma$& Cross sectional area\\
 $P$& Pressure\\
 $\eta$& Viscosity of liquid\\
 $r$& Radius of nanofluidic pore\\
 $L$& Flow length of instability\\
 $\alpha_{e}$& Element size in the simulation\\
 $N_{e}$& Number of elements in simulation\\
 $\Phi$& Area of nanopore\\
 $\bm{r}$& Position vector\\
 $b$& Distance constant\\
 $\bm{\beta}$& Constant velocity component in a vortex pair\\
 $(u, v, w)$& Velocity components in $x, y, z$ direction\\
 $R_e$& Reynolds number\\
 $E$& Energy\\
 $k_{eff}$& Effective thermal conductivity\\
 suffix $_{g, l}$& Gas phase and liquid phase, respectively\\
 suffix $_{a, s}$& Analytical and simulation, respectively\\
 suffix $_{in}$& Interface\\
\hline
\end{tabular}
\end{table}

\section{Governing differential equations}
The continuity equation is given by:
\begin{equation}
   \frac{\partial{\rho}}{\partial{t}} +  \frac{\partial{\rho\bm{u}}}{\partial{x}} + \frac{\partial{\rho\bm{v}}}{\partial{y}} + \frac{\partial{\rho\bm{w}}}{\partial{z}} = 0 
   \label{A1}
\end{equation}
where $(u, v, w)$ are the velocity components in $(x, y, z)$ direction respectively, $t$ is the time (usually a large time period is considered for steady state simulation), and $\rho$ is the density.
This is a fundamental equation in fluid dynamics and represents the conservation of mass for a fluid. In this equation:\\
- \( \frac{\partial{\rho}}{\partial{t}} \) represents the rate of change of density (\( \rho \)) with respect to time (\( t \)).\\
- \( \frac{\partial{\rho\bm{u}}}{\partial{x}} \) represents the rate of change of the density (\( \rho \)) times the velocity component (\( \bm{u} \)) in the x-direction (\( x \)) with respect to \( x \).\\
- \( \frac{\partial{\rho\bm{v}}}{\partial{y}} \) represents the rate of change of the density (\( \rho \)) times the velocity component (\( \bm{v} \)) in the y-direction (\( y \)) with respect to \( y \).\\
- \( \frac{\partial{\rho\bm{w}}}{\partial{z}} \) represents the rate of change of the density (\( \rho \)) times the velocity component (\( \bm{w} \)) in the z-direction (\( z \)) with respect to \( z \).\\
The equation states that the total rate of change of density within a fluid element is equal to zero, which means that mass is conserved. In simpler terms, it asserts that the change in density at a given point in space and time is equal to the net flow of mass into or out of that point.
This equation is a fundamental part of the Navier-Stokes equations, which describe the behaviour of fluid flow.
The momentum equation in equation \ref{A1} is given by\\
$x$-component 
\begin{equation}
\begin{split}
\frac{\partial {(\rho \bm{u})}}{\partial t} + {\frac{\partial{(\rho \bm{u}^2})}{\partial{x}}} + {\frac{\partial{(\rho \bm{u} \bm{v}})}{\partial{y}}} + {\frac{\partial{(\rho \bm{u} \bm{w}})}{\partial{z}}} = \\ {-\frac{\partial{p}}{\partial{x}}} + {\frac{1}{Re}}
\bigg[
{\frac{\partial{\rttensor\tau_{xx}}}{\partial{x}}} + {\frac{\partial{\rttensor\tau_{xy}}}{\partial{y}}}+{\frac{\partial{\rttensor\tau_{xz}}}{\partial{z}}}
\bigg]
\end{split}
\end{equation}

\noindent $y$-component
\begin{equation}
\begin{split}
\frac{\partial {(\rho \bm{v})}}{\partial t} + {\frac{\partial{(\rho \bm{u}\bm{v}})}{\partial{x}}} + {\frac{\partial{(\rho \bm{v}^2})}{\partial{y}}} + {\frac{\partial{(\rho \bm{v} \bm{w}})}{\partial{z}}} = \\ {-\frac{\partial{p}}{\partial{y}}} + {\frac{1}{Re}} \bigg[{\frac{\partial{\rttensor\tau_{xy}}}{\partial{x}}} + {\frac{\partial{\rttensor\tau_{yy}}}{\partial{y}}}+{\frac{\partial{\rttensor\tau_{yz}}}{\partial{z}}}\bigg]
\end{split}
\end{equation}
\noindent $z$-component
\begin{equation}
\begin{split}
\frac{\partial(\rho \bm{w})}{\partial t} + {\frac{\partial{(\rho \bm{u}\bm{w})}}{\partial{x}}} + {\frac{\partial{(\rho \bm{v}\bm{w})}}{\partial{y}}} + {\frac{\partial{(\rho \bm{w}^2)}}{\partial{z}}} =\\ {-\frac{\partial{p}}{\partial{z}}} + {\frac{1}{Re}} \bigg[{\frac{\partial{\rttensor\tau_{xz}}}{\partial{x}}} + {\frac{\partial{\rttensor\tau_{yz}}}{\partial{y}}}+{\frac{\partial{\rttensor\tau_{zz}}}{\partial{z}}}\bigg]
\end{split}
\end{equation}

\noindent The energy equation is given by
\begin{equation}
    \frac{\partial(\rho E)}{\partial t} + \nabla.(\bm{v}(\rho E + P)) = \nabla.(k_{eff} \Delta T)
\end{equation}
where $E$ represents the internal energy, $P$ is the pressure, and $T$ is the temperature, which are mass-averaged variables according to the volume fraction model. The phase parameter $p$ is included to account for multiphase effects, and $k_{eff}$ denotes the effective thermal conductivity.
\cite{manual2009ansys}.


\begin{figure*}[h!]
\centering
\includegraphics[width=0.8\textwidth]{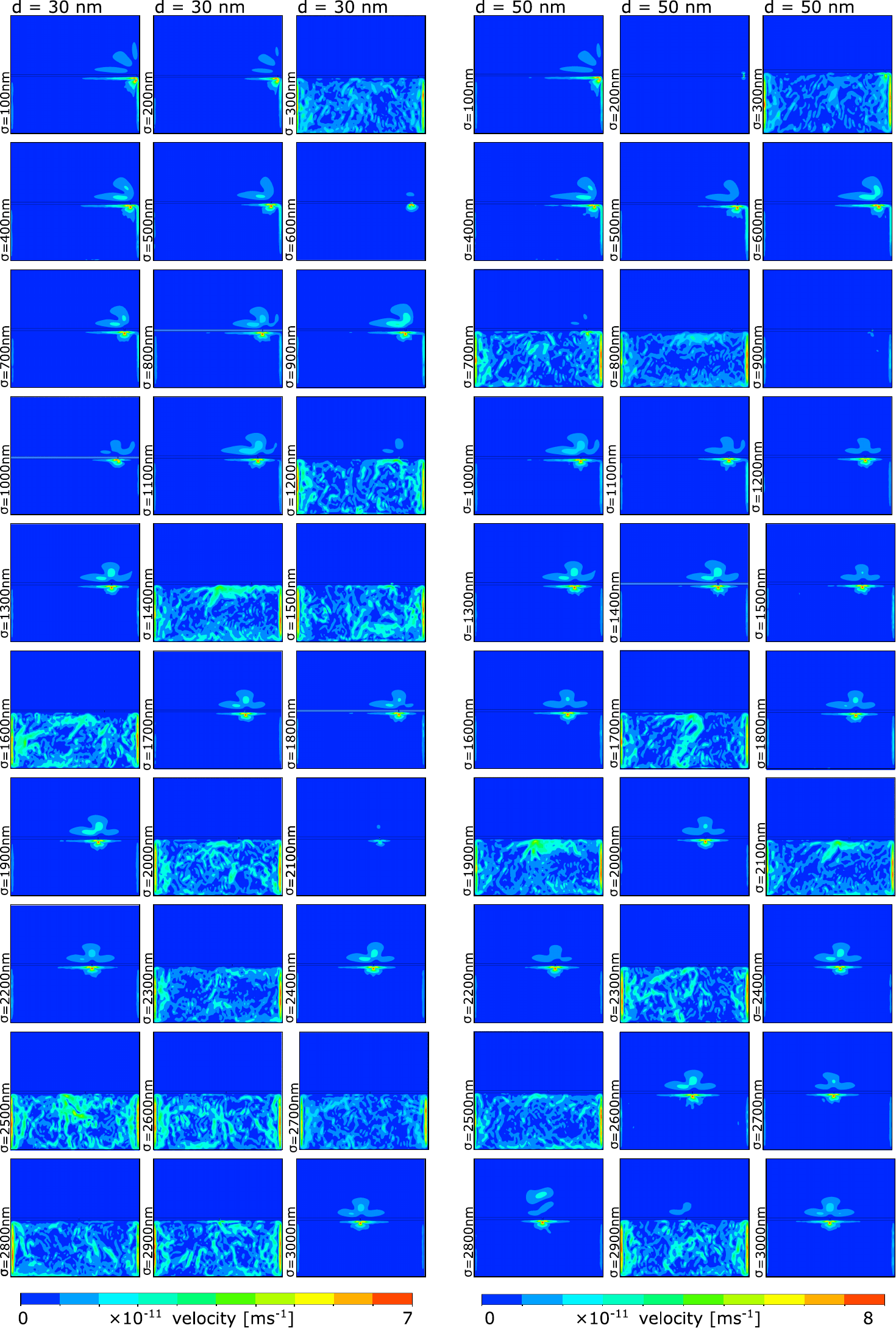}
\caption{Detailed flow dynamics of 30 nm and 50 nm single nanofluidic pore systems.
         }
\label{Figure:6}
\end{figure*}

\newpage
\begin{figure*}[h!]
\centering
\includegraphics[width=0.4\textwidth]{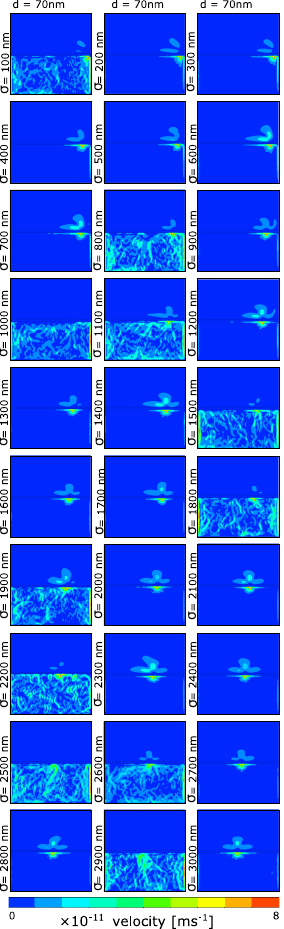}
\caption{Detailed flow dynamics of 70 nm single nanofluidic pore system.}
\label{Figure:7}
\end{figure*}

\begin{figure*}
\includegraphics[width=1\textwidth]{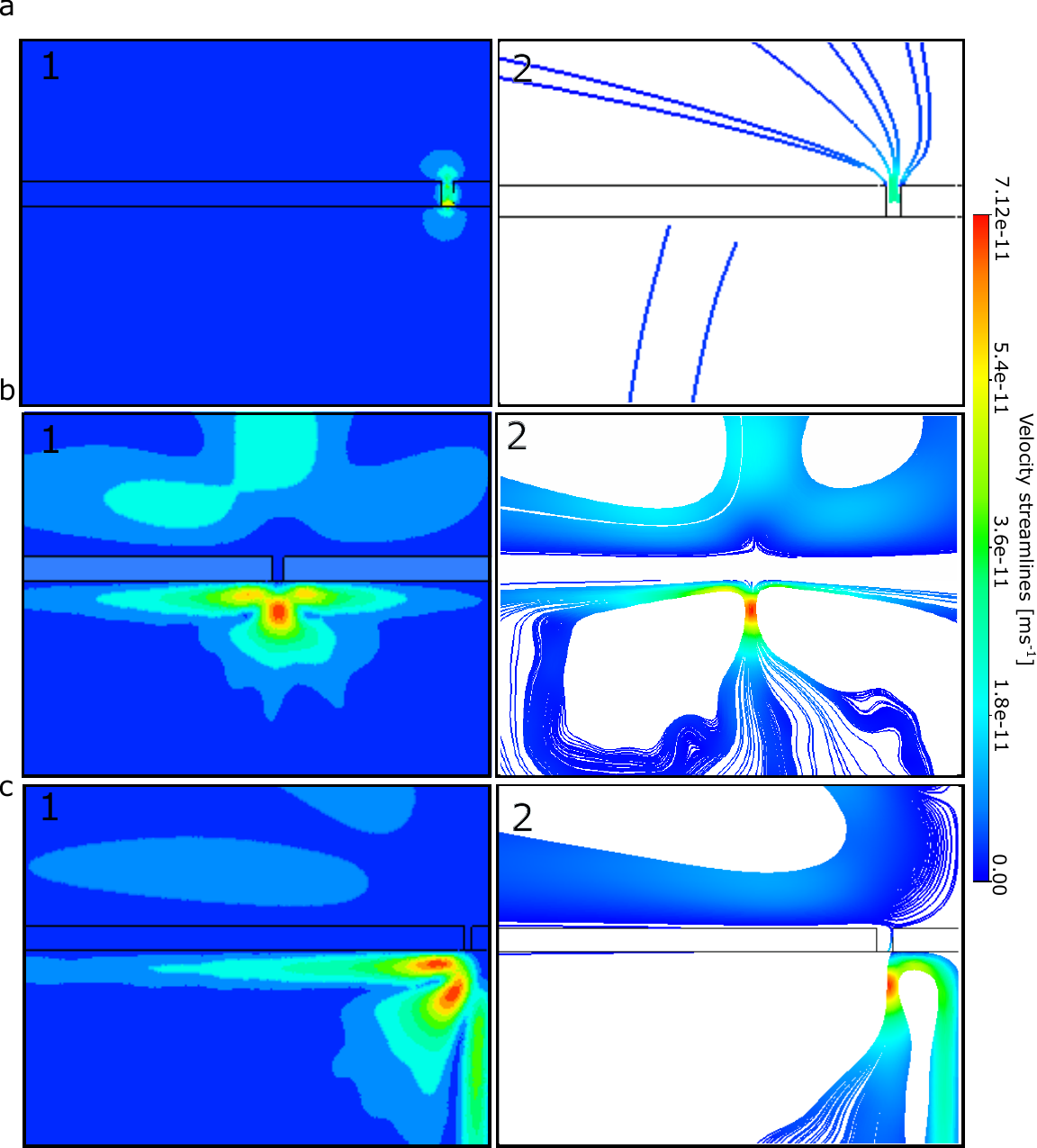}
\caption{(a), (b), (c) Left: Major velocity magnitude distribution patterns and right: velocity streamlines. }
\label{Figure:8}
\end{figure*}

\begin{figure*}
\includegraphics[width=0.6\textwidth]{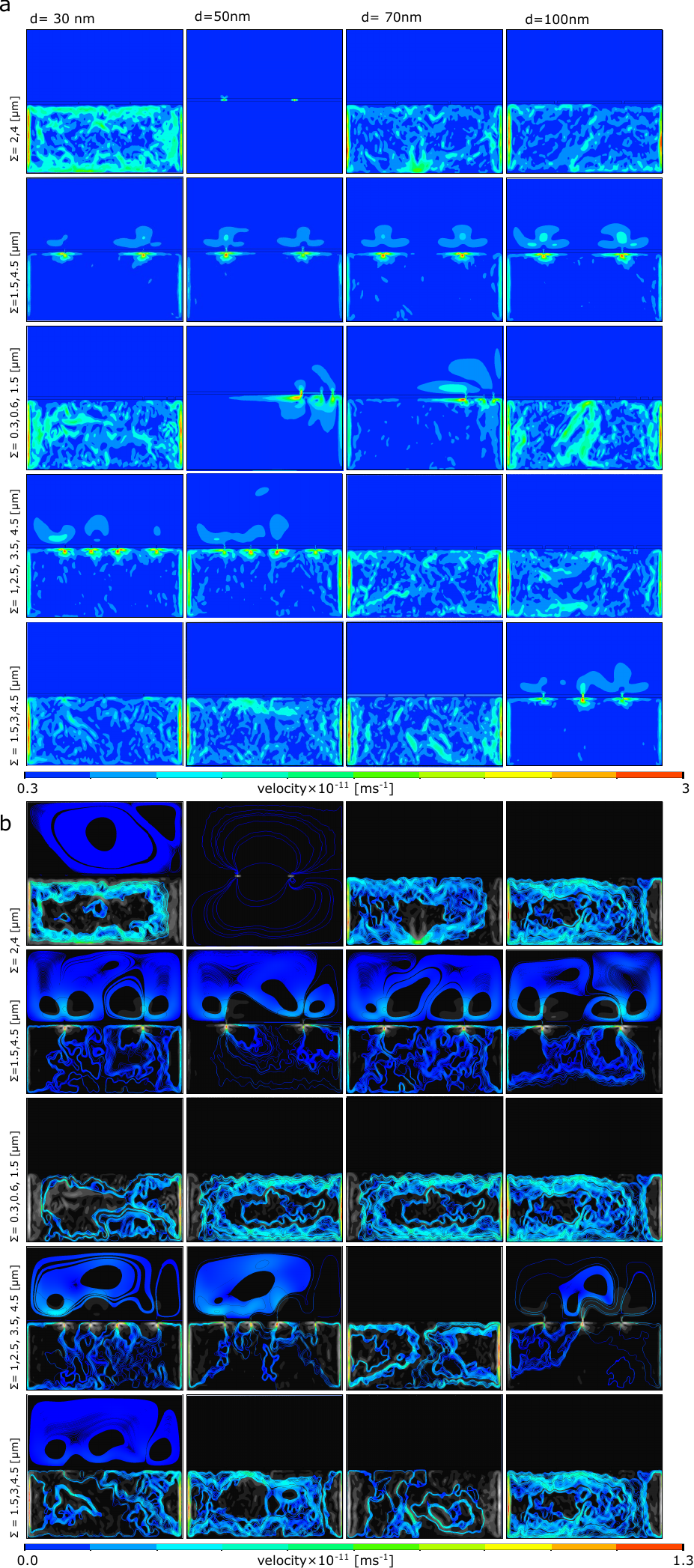}
\caption{(a) Velocity magnitude distribution patterns and (b) velocity streamline distribution. }
\label{Figure:9}
\end{figure*}

\begin{figure*}
\includegraphics[width=1\textwidth]{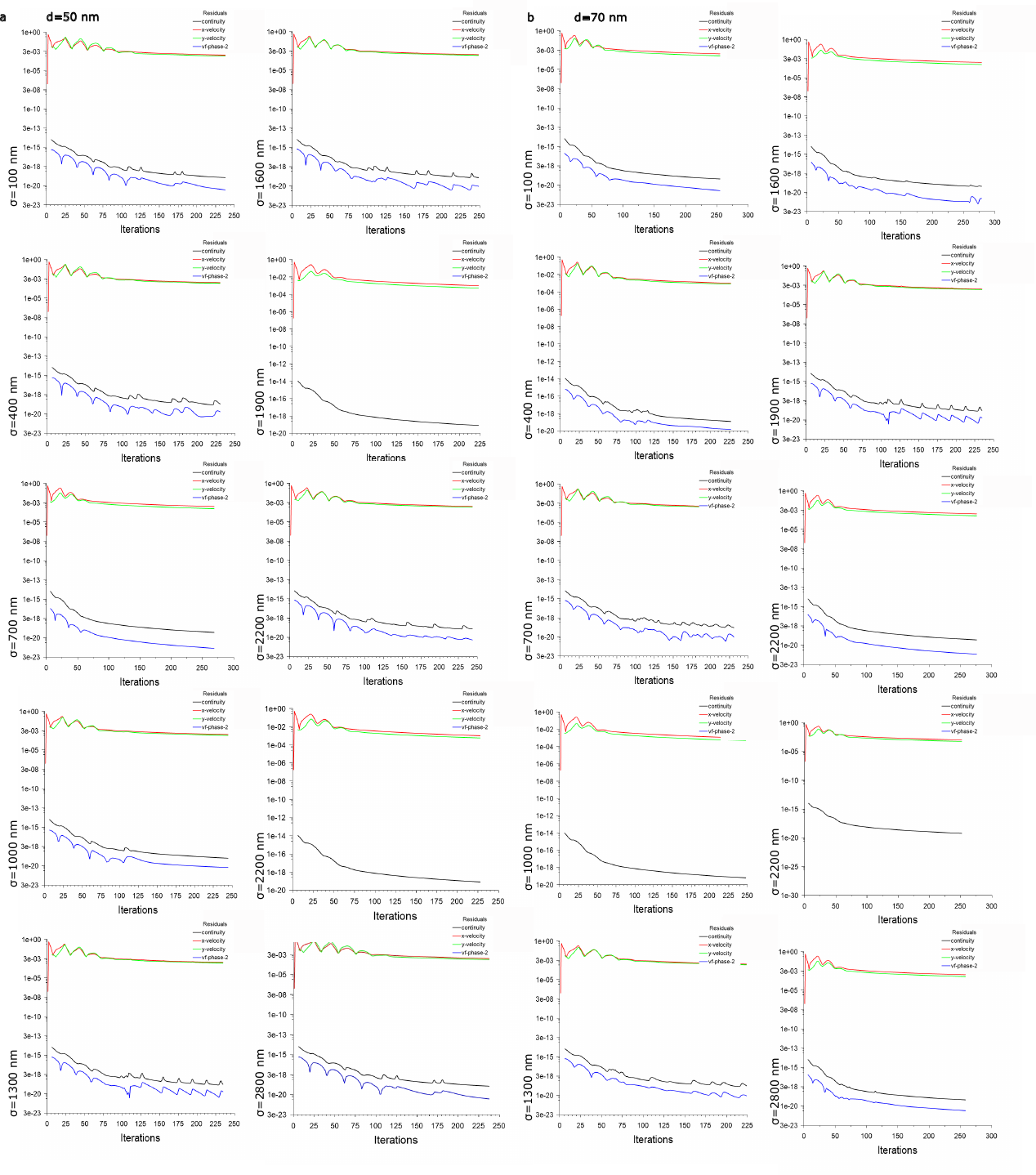}
\caption{absolute convergence data of residuals within the nanopore systems. (a) 50 nm pore size singgle nanopore systems, (b) 70 nm pore size single nanopore systems. }
\label{Figure:10}
\end{figure*}

\begin{figure*}
\includegraphics[width=1\textwidth]{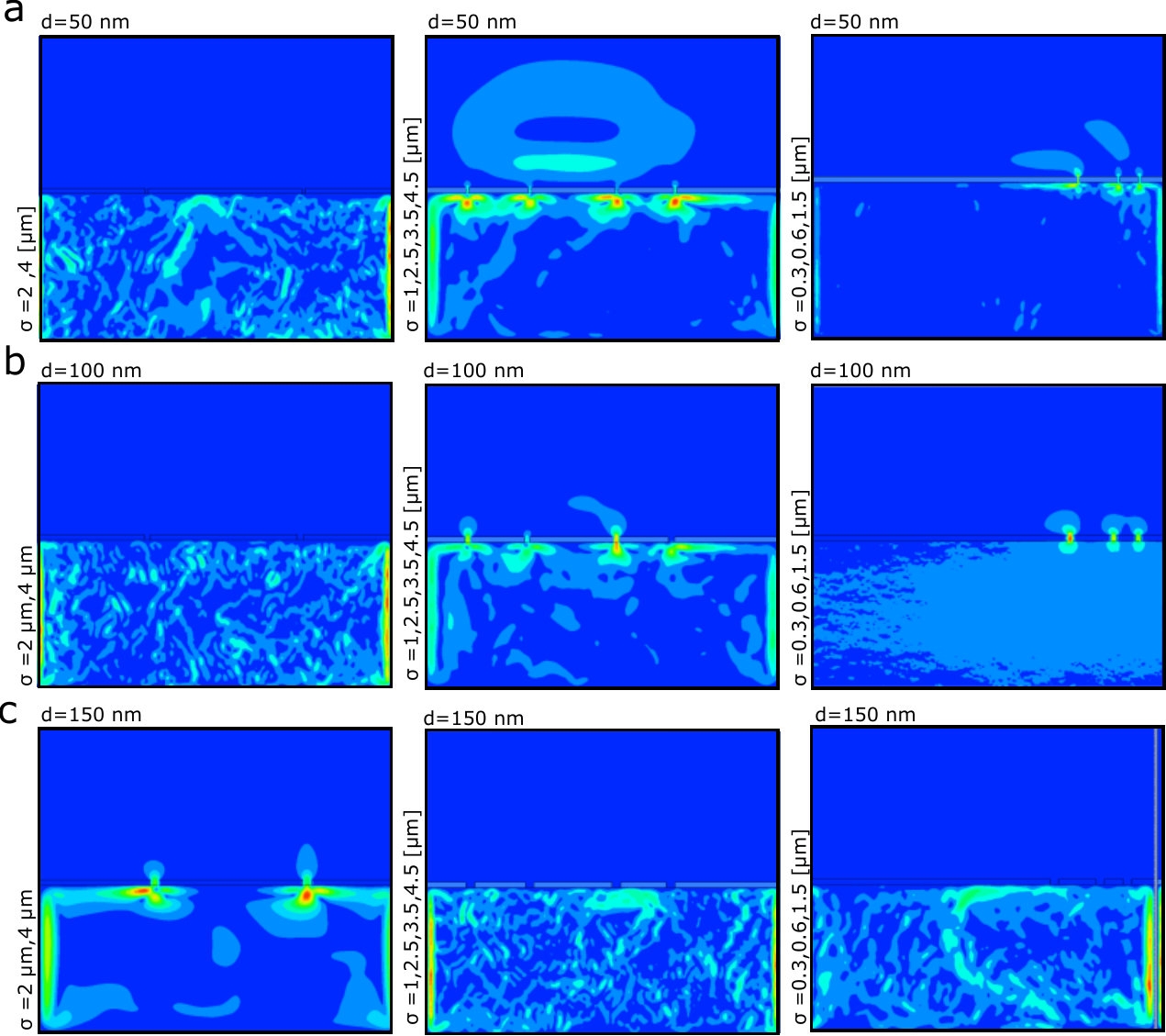}
\caption{Flow dynamics through the nanopores at acetone-air interface. (a) 50 nm nanopore systems with varying $\sigma$, (b) 70 nm nanopore systems with varying $\sigma$, (c) 100 nm nanopore systems with varying $\sigma$. }
\label{Figure:11}
\end{figure*}

\end{document}